\newif\ifShowKeys
\ShowKeystrue
\ShowKeysfalse

\newif\ifshowtikz
\showtikztrue
\showtikzfalse   % <---- comment/uncomment that line

\documentclass[11pt]{article}
	\pdfoutput=1
\usepackage[T1]{fontenc}

\usepackage{listings}
\lstnewenvironment{arkady}  % usage \begin{arkady} ... \end{arkady}
{\lstset{language=C,frame=trbl,basicstyle = \footnotesize \ttfamily , breaklines = true,showstringspaces=false}}{}

\usepackage{froufrou}
\usepackage{datetime}
\usepackage{comment}					% to comment large parts of text

\usepackage[usenames,dvipsnames]{xcolor}
\usepackage[setpagesize=false,pagebackref=false, 
linktocpage, bookmarksopen=true, colorlinks=true, 
linkcolor=Maroon,citecolor=Maroon,urlcolor=Maroon]{hyperref}

\usepackage[parsep]{collref}				% collect references in groups

\ifShowKeys \usepackage[notcite]{showkeys} \fi

\usepackage{amsmath, amssymb,amsthm}
\usepackage{stackrel}
\numberwithin{equation}{section}
\usepackage{bm,environ,mathrsfs,array,arydshln}
\usepackage{booktabs,float,slashed}
\usepackage{appendix}
\usepackage[mathcal]{euscript}
\usepackage{tensor} 						% Ratcliffe package to write tensors
\usepackage{mathabx}
\usepackage[vcentermath]{youngtab}
\usepackage{simpler-wick}

\usepackage{graphicx,epsfig,epic}
\usepackage{subcaption,wrapfig}
\usepackage{tikz}
\usepackage{tikz-feynman} 
\allowdisplaybreaks

\usepackage{framed}						% for shaded equations \begin{shaded} + \end{shaded} or \bs + \es
\definecolor{shadecolor}{rgb}{0.9996078, 0.984314, 0.960784}
\definecolor{framecolor}{rgb}{0,0,0}
\definecolor{TFTitleColor}{RGB}{1,1,1}
\definecolor{TFFrameColor}{RGB}{249	218	181}		
\definecolor{TFFrameColor}{RGB}{230 230 230 }

\newenvironment{frshaded}{%
    \MakeFramed {\FrameRestore}}%
    {\endMakeFramed}

\definecolor{myred}{RGB}{233, 33, 45}

\usepackage{mdframed}

\definecolor{lightpeach}{RGB}{255, 247, 235}

\newmdenv[
  backgroundcolor=gray!5,
  linecolor=gray!40,
  roundcorner=5pt,
  innerleftmargin=8pt,
  innerrightmargin=8pt,
  innertopmargin=6pt,
  innerbottommargin=6pt,
]{remarkbox}

\newenvironment{lab-remark}[1]{
  \begin{remarkbox}
  \textbf{\raisebox{0.2ex}{$\triangledown$} \textbf{ #1 } }\ignorespaces
}{
  \end{remarkbox}
}

\newcommand{\bs}{\begin{frshaded}}			% framed with background in shadecolor 
\newcommand{\es}{\end{frshaded}\noindent}

\def\ba#1\ea{\begin{align}#1\end{align}}		        %  clever way to bypass the known problem...
\newcommand{\be}{\begin{equation}}
\newcommand{\ee}{\end{equation}}
\newcommand{\bea}{\begin{equation} \begin{aligned}} 
\newcommand{\eea}{\end{aligned} \end{equation}}
\newcommand{\mc}{\mathcal }

\newcommand{\wt}{\widetilde}

\newcommand{\mk}{\mathfrak}
\newcommand{\la}{\label}
\newcommand{\eps}{\varepsilon}

\newcommand{\lp}{\notag \\ & }

\DeclareMathOperator{\sech}{\text{sech}}

\newcommand{\cf}{\textit{cf.} }
\newcommand{\ie}{\textit{i.e.} }

\renewcommand{\l}{\lambda}

\newcommand{\ket}[1]{|#1\rangle}
\newcommand{\bra}[1]{\langle #1|}
\newcommand{\braket}[2]{\langle #1|#2\rangle}
\newcommand{\mmm}[3]{\langle #1|#2|#3\rangle}

\newcommand{\UV}{\eps_{\scalebox{0.6}{UV}}}

\begin{document}

\begin{comment}
\centerline{\Large\sc  Krylov Veneziano--Wosiek notes}
\vskip 0.2cm
\centerline{-- Notes --}
\vskip 0.2cm\centerline{\small\today\ -- \currenttime}
\vskip 0.5cm
 %\centerline{\sc M. Beccaria}
\bigskip
\begin{abstract}
\begin{center}
%\includegraphics[width=0.5\textwidth]{cover}
\end{center}
\end{abstract}
\end{comment}

%%%%%%%%%%%%%%%%%%%%%%%%%%%%%%%%%%%%%%%%%%%%%%%%%%%%%%%%%
%%%%% NOTICE the title page is commented out with \begin{comment}...\end{comment}   but is ready to be used

%\begin{comment}

\begin{titlepage}
%\begin{tabbing}
%\hspace*{11.5cm} \=  \kill % set the tabbings
%\>  Imperial-TP-AT-2024-?? \\
%\> %none
%\end{tabbing}

%\centerline{\small\today\ -- \currenttime}

\vspace*{15mm}
\begin{center}
{\Large  Krylov Correlators in $\mk{sl}(2,\mathbb R)$ Models: \\ \vskip 5pt 
Exact Results and Holographic Complexity} %\vskip 9pt
%{\Large\sc      Notes}
\vspace*{10mm}

E. Alfinito$^{a}$ and M. Beccaria$^{a,b}$

\vspace*{4mm}
{\small
	
${}^a$ Universit\`a del Salento, Dipartimento di Matematica e Fisica \textit{Ennio De Giorgi},
\vskip 0.2cm
${}^{b}$ INFN - sezione di Lecce, Via Arnesano, I-73100 Lecce, Italy
\vskip 0.3cm
\vskip 0.2cm {\small E-mail: \texttt{matteo.beccaria@le.infn.it}}
}
\vspace*{0.8cm}
\end{center}

\begin{abstract}  
In holography, the complexity--momentum correspondence relates the increasing momentum 
of a point particle falling into an eternal black hole to the rate of growth of the 
Krylov complexity of the dual boundary state, a conjecture established exactly for the 
BTZ black hole in AdS$_{3}$ at the semiclassical level. We examine possible extensions 
of the correspondence by considering boundary higher Krylov complexities and Krylov 
correlators encoding fluctuations and temporal correlations of the spreading quantum state.
To this end, we derive exact results for Krylov correlators in quantum systems 
with $\mathfrak{sl}(2,\mathbb{R})$ or Heisenberg-Weyl symmetry and apply them 
to the complexity--momentum correspondence. 
We show that certain out-of-time-ordered correlators of two or more Krylov speed operators
at different times 
are proportional to combinations of the proper radial momenta of a particle 
falling into the BTZ black hole in AdS$_{3}$, evaluated at those times.
This represents a first step 
in the generalization of the original complexity--momentum relation.
\end{abstract}
\end{titlepage}

%\end{comment}
%%%%%%%%%%%%%%%%%%%%%%%%%%%%%%%%%%%%%%%%%%%%%%%%%%%%%%%%%

{\small
\makeatletter
\newcommand*{\toccontents}{\@starttoc{toc}}
\makeatother
\toccontents
}

%\tableofcontents

\vspace{1cm}

\setcounter{footnote}{0}

\section{Introduction}
\label{sec:intro}

Among the various notions of complexity at the intersection of quantum information,
many-body physics, and quantum gravity, Krylov complexity has emerged as a
particularly natural and computable measure of quantum information spreading
under Hamiltonian evolution   \cite{Nandy:2024evd,Baiguera:2025dkc,Rabinovici:2025otw}.
The concept grew out of the study of out-of-time-order correlators 
(OTOCs), which capture early-time chaos through a Lyapunov exponent 
$\lambda_{\rm L}$ bounded by $2\pi k_{\rm B}T/\hbar$ 
\cite{Maldacena:2015waa}, but do not directly resolve the 
\textit{structure} of operator spreading in Hilbert space.
The seminal work \cite{Parker:2018yvk} addressed this by introducing Krylov
complexity: applying the Lanczos algorithm to the operator $\mathcal{O}$ under
the Liouvillian $[H,\cdot]$ generates an orthonormal basis
$\{|\mathcal{O}_{n}\rangle\}$, and Krylov complexity is the average position of
$\mathcal{O}(t)$ along this chain. The \textit{universal operator growth hypothesis}
of \cite{Parker:2018yvk} states that in chaotic systems the Lanczos coefficients
$b_n$ grow linearly, $b_n \sim \alpha n$, with $\lambda_{\rm L}\le 2\alpha$,
making Krylov complexity a finer probe of chaos than the OTOC alone.
Krylov complexity has since found applications in many-body physics and quantum
field theory \cite{Parker:2018yvk,Barbon:2019wsy,Rabinovici:2021qqt,Rabinovici:2022beu}
and in holography, 
where it has been studied in various contexts, with a notable example being its identification 
with the wormhole length in SYK/JT duality  \cite{Rabinovici:2023yex,Xu:2024gfm,Ambrosini:2024sre,Heller:2024ldz}.
%\cite{Rabinovici:2023yex,Hashimoto:2026kjy,Roychowdhury:2026sgg}.

A central concept in holographic complexity is the \textit{eternal black hole}
\cite{Maldacena:2001kr}, whose boundary dual is the thermofield double (TFD) state:
a purification of the thermal density matrix in two copies of the CFT Hilbert space.
Its holographic dual is the BTZ black hole in AdS$_3$, whose interior
Einstein--Rosen bridge grows monotonically (linearly at late times) --- one of the central features of Susskind's
Complexity$=$Volume conjecture (C=V) \cite{Susskind:2014rva,Stanford:2014jda,Brown:2015lvg}, identifying the wormhole volume with the
complexity of the dual state. This picture was later generalised in \cite{Belin:2021bga,Belin:2022xmt}
to an infinite family of 
boundary-anchored codimension-one bulk surfaces, each of which grows linearly 
at late times --- the ``complexity=anything'' proposal.

A related and more specific conjecture
\cite{Susskind:2014jwa,Susskind:2018tei,Susskind:2019ddc,Susskind:2020gnl,Lin:2019qwu}
states that the \textit{rate} of growth of the C=V holographic complexity $\mathcal{C}(t)$
is proportional to the proper radial momentum of a massive infalling particle,
\be
\la{1.1}
\mathcal{C}'(t) \,\propto\, P_{\rho}\,.
\ee
This was made precise in \cite{Barbon:2019tuq,Barbon:2020olv,Barbon:2020uux}, which
proved that $d\mathcal{C}/dt$ equals the
integrated matter momentum flux through the maximal bulk slice, exactly
in $2+1$ dimensions and for spherically
symmetric solutions in arbitrary dimensions.

The main obstacle to establishing a relation of the type (\ref{1.1}) 
with an intrinsic \textit{boundary} definition on the left-hand side 
was the absence of a microscopically well-defined notion of 
complexity on the CFT side.
In the remarkable paper \cite{Caputa:2024sux} this obstacle was overcome 
by replacing $\mathcal{C}(t)$ with the spread (Krylov) complexity $C(t)$ as the
boundary definition, 
\footnote{It is important to distinguish \textit{operator} Krylov complexity 
\cite{Parker:2018yvk}, defined in the Heisenberg picture via the Lanczos 
algorithm on operator space, from \textit{state} (or spread) complexity 
\cite{Balasubramanian:2022tpr}, defined in the Schr\"odinger picture for 
time-evolved states. In this paper we work exclusively in the state 
complexity framework.}
with the initial 
state being the TFD excited by a local primary operator of conformal dimension 
$\Delta$, identified with the mass $m=\Delta$ of the dual infalling particle.  
With this choice, and with the crucial observation that the
relevant bulk radial coordinate is the \textit{proper} distance $\rho$ from the
horizon (defined by $z = e^{-\rho}$ in Poincar\'e AdS, or its analogue in the black
hole background), the correspondence becomes a genuine holographic duality,
with the left-hand side computed entirely from the boundary CFT
and the right-hand side from classical bulk geodesics:
\be
\la{1.2}
C'(t) \, \propto\, P_{\rho}\,.
\ee
This constitutes the first CFT-derived 
instance of the momentum--complexity correspondence, promoting it from a qualitative
conjecture to a new entry in the holographic dictionary for the duality AdS$_3$/CFT$_2$. \footnote{
The correspondence is extended to smeared (non-local) operators 
and to $d>2$ dimensions in \cite{Aguilar-Gutierrez:2025kmw}, using 
Krylov operator complexity in Rindler-AdS$_{d+1}$, with exact equality 
holding in the asymptotic boundary and near-horizon limits.}
Subsequent developments of the correspondence have been largely pursued on 
the bulk side --- extending it to higher dimensions, charged probes, and 
non-conformal backgrounds \cite{Fan:2024iop,He:2024pox,Fatemiabhari:2025cyy,
Fatemiabhari:2025usn} --- while the boundary CFT perspective, initiated in 
\cite{Caputa:2024sux}, has remained comparatively unexplored. The present 
paper is a step in the opposite direction.

The result of \cite{Caputa:2024sux} is the starting point of our 
investigation, and we now delineate precisely what it 
establishes and what it leaves open.
The computation relies on the semi-classical 
identification of the operator insertion with a point particle of mass 
$m = \Delta$, which requires $\Delta\gg 1$. This allows one to treat the probe as a classical 
particle moving on a geodesic of the BTZ background, with $1/\Delta$ 
corrections corresponding to quantum fluctuations 
around the classical trajectory. The background geometry remains 
classical throughout. 

Remarkably, the spread complexity is computed using 
only the two-point function of $\mc O$, which is fixed by conformal symmetry 
alone and is independent of the central charge $c$. In particular, the BTZ 
black hole's  quantum 
gravitational properties --- controlled by the  central charge $c = 3L_{\rm AdS}/(2G_N)$  of the boundary CFT --- are invisible 
to the correspondence at this level. This is not surprising since a classical bulk geometry 
requires $c\gg 1$ in the first place.

This raises a question about the depth of the result: the correspondence 
holds in a regime where the bulk geometry is purely classical and the CFT 
calculation requires only conformal kinematics, without probing any 
genuinely quantum gravitational physics. Whether it survives in regimes 
where quantum gravity effects are relevant --- finite $c$, off-shell 
geometries, stringy corrections --- remains to be seen.
In fact, already at the level of the 
semiclassical result, the spread complexity in the present setup 
fails to capture two physical features that are expected on general 
grounds for a quantum black hole. The first is  saturation of complexity at 
times of order $e^{S}$ (where $S$ is the Bekenstein-Hawking entropy), 
due to quantum recurrences in the finite-dimensional Hilbert space 
\cite{Brown:2017jil,Susskind:2018pmk}. The second is the  switchback effect 
\cite{Stanford:2014jda,Brown:2015lvg,Ambrosini:2025hvo,Aguilar-Gutierrez:2025mxf}, whereby a perturbation 
inserted at early times reduces the complexity growth rate by an 
amount controlled by the scrambling time. Both effects require 
physics beyond the classical probe approximation used here.
%A further  limitation, of a more technical nature,  is that the underlying
%$\mk{sl}(2,\mathbb R)$ symmetry that makes the spread complexity analytically tractable
%is special to two-dimensional CFTs. 

To make these statements precise and set up the framework 
of this paper, we recall in more detail the definition of \textit{spread} Krylov complexity.
On the CFT side, the Hamiltonian evolution of any quantum state
is described as the propagation along a chain of states built according to the Lanczos algorithm.
Given an initial state $\ket{K_{0}}$ in a system with Hamiltonian $H$, the Lanczos
algorithm generates iteratively an orthonormal basis $\{\ket{K_{n}}\}_{n\ge 0}$
of the subspace of Hilbert space reachable from $\ket{K_{0}}$ by repeated action of
$H$. By construction, $H$ acts tridiagonally in this basis, with diagonal 
coefficients $a_n$ and off-diagonal coefficients $b_n$, both real, encoding 
all dynamical information.
The Krylov number operator $N$, defined by $N\ket{K_{n}} = n\ket{K_{n}}$, measures
position along this chain. The spread complexity is then simply
\be
C(t) = \mmm{K_{0}}{N(t)}{K_{0}},
\ee
where $N(t) = e^{itH}Ne^{-itH}$ is the
Heisenberg-evolved operator. The complexity rate of growth $C'(t)$ entering the complexity--momentum correspondence 
is  the one-point function of a Krylov speed operator $N'(t) = i[H,N(t)]$. 

A natural question is  whether Krylov correlators (of products of $N(t)$ and its derivatives at different times) also have a bulk interpretation. 
The aim of this paper is to start an investigation of this issue by 
looking at a natural yet largely unexplored extension: rather than studying the single one-point function $C(t) $
that defines the spread complexity, we study the multi-point correlators of the Heisenberg-evolved Krylov number operator
\be
\la{1.4}
C_{p}(t_{1},\dots,t_{p}) = \mmm{K_{0}}{N(t_{1})\cdots N(t_{p})}{K_{0}},
\ee
and the specialized equal-time higher complexities --- or Krylov moments ---
\be
\la{1.5}
C^{(p)}(t) \equiv C_{p}(t, \dots, t) =  \mmm{K_{0}}{N(t)^{p}}{K_{0}}.
\ee
%and the fully antisymmetrized $p$-point correlators
%\be
%\la{1.6}
%A_{p}(t_1,\dots,t_p)
%= \sum_{\sigma\in S_p}(-1)^\sigma
%\mmm{K_0}{N(t_{\sigma_1})\cdots N(t_{\sigma_p})}{K_0},
%\ee
%starting, for $p=2$,  with the commutator expectation value $A_{2}(t_1, t_{2}) = \mmm{K_0}{[N(t_{1}), N(t_{2})]}{K_0}$.
Unlike standard time-ordered Green's functions, the 
correlators (\ref{1.4}) preserve the operator ordering, 
so correlators involving $N' = dN/dt$ follow by differentiating 
in the insertion times. We focus on $N'$ because its one-point 
function $C'(t)$ is the quantity entering the 
complexity--momentum correspondence (\ref{1.2}); higher 
derivatives are in principle accessible but have no known 
bulk interpretation at this stage.

These objects encode the full statistics of the position distribution
along the Krylov chain and interpolate between the standard Krylov complexity
(at $p=1$) and 
OTOC-like quantities in Krylov space, \ie
specific linear combinations of (\ref{1.4}) that are sensitive to the non-commutativity of $N(t)$ at different times.
The higher moments $C^{(p)}$ provide
finer information about the spread of the wave function in the Krylov basis.
For instance, the variance $\text{Var}(t) = C^{(2)}(t) - C(t)^2$
measures the width of the distribution 
\cite{Bhattacharjee:2022lzy,Bhattacharjee:2022ave} and 
higher cumulants  
have been considered in \cite{Bhattacharjee:2022ave,Fu:2024fdm,Camargo:2024rrj,Grabarits:2025xys},
though without the exact closed-form results we derive here for 
$\mathfrak{sl}(2,\mathbb{R})$ and Heisenberg models, 
nor their holographic interpretation.

%The 
%antisymmetrized multi-point correlators
%capture genuinely non-commutative aspects of Krylov
%dynamics.

In order to compute Krylov correlators explicitly, we work in a class of quantum systems, 
identified early on in \cite{Caputa:2021sib,Balasubramanian:2022tpr}, whose spread 
complexity is analytically tractable.
In these models, the Hamiltonian is built from the generators of
$\mk{sl}(2,\mathbb R)$ or, in a suitable limit, of the Heisenberg-Weyl algebra. \footnote{See \cite{Grabarits:2026hjz} for 
 exact results for arbitrary time-dependent generators.}
The Krylov basis coincides with the representation-theoretic
basis of the relevant Lie algebra, and spread complexity reduces to a matrix
element of a generator. This observation has been exploited in several papers to compute $C(t)$ in closed form, see in particular
\cite{Caputa:2021sib,Balasubramanian:2022tpr,Rabinovici:2023yex,Caputa:2024sux}.
Extension to Krylov correlators 
exploits this structure, \ie the 
observation that the Heisenberg equation of motion for the Krylov number
operator $N(t)$ closes exactly within the algebra, so that any multi-point
correlator can be expressed in terms of  algebraically computable matrix elements. 

The general formulas that we derive have a direct application to the complexity--momentum correspondence.
Indeed, it is known that the Krylov coefficients for the chain starting with the TFD state 
excited by a local primary operator match those of a Hamiltonian with $\mk{sl}(2,\mathbb R)$ symmetry,
as follows from conformal symmetry \cite{Caputa:2024sux,Caputa:2023vyr}.  It is thus possible to look for a bulk interpretation
of the  Krylov correlators  in terms of proper momentum. 

Our results show that this is indeed possible for specific 
partially antisymmetrized combinations of $N'(t)$ at different times.
In particular, we find that  the commutator of two Krylov speed operators is 
proportional to the proper momentum difference at the two times $t_{1}$, $t_{2}$. 
Concretely, denoting by $P_{i}$ the proper (radial) momentum at time $t_{i}$, and $\langle\mc O\rangle\equiv \mmm{K_{0}}{\mc O}{K_{0}}$, we find
\be
\la{1.7}
i\,\langle [N'(t_{1}), N'(t_{2})] \rangle = -\frac{P_{1}-P_{2}}{\UV^{2}}+O(1/\Delta^{2}),
\ee
where the UV cutoff $\UV$ sets the particle's initial position in the bulk, near the AdS boundary. Similarly, by considering Krylov correlators of three 
speed operators at different times, we find the cubic relation
\be
\la{1.8}
i\,\langle N'(t_{1})\,N'(t_{2})\,N'(t_{3})-N'(t_{3})\,N'(t_{2})\, N'(t_{1})\rangle = -2\, \frac{1}{\UV^{3}}\, P_{2}\, (P_{1}-P_{3})+O(1/\Delta),
\ee
while with four operators one has for instance
\be
\la{1.9}
\langle [N'(t_{1}), N'(t_{2})]\, [N'(t_{3}), N'(t_{4})] \rangle = -\frac{1}{\UV^{4}}\, (P_{1}-P_{2})\, (P_{3}-P_{4})+O(1/\Delta).
\ee
The analysis of higher moments (\ref{1.5}) (and their counterparts with $N\to N'$) is instead
less satisfactory since they are too degenerate in the semiclassical limit to extend the complexity--momentum correspondence at leading order.
In general, our exact formulas also provide explicit $1/\Delta$ corrections to 
(\ref{1.7})--(\ref{1.9}), which constitute  simple CFT predictions 
for quantum corrections to the probe trajectory in the BTZ background, 
testable  by worldline quantum mechanics of the dual bulk field.

To our knowledge, Eqs.~(\ref{1.7}), (\ref{1.8}), and (\ref{1.9})  constitute the first instances of a many-time Krylov observable admitting a precise semiclassical bulk interpretation, 
extending, albeit in a minor way,  the complexity--momentum correspondence from a one-point to a genuinely multi-time statement.

\paragraph{Structure of the paper.}
Section~\ref{sec:general} reviews general facts about  Krylov complexity.
Section~\ref{sec:math} develops the theory
of Krylov correlators. In Section \ref{sec:sl2} we discuss $\mk{sl}(2,\mathbb R)$ models, deriving closed-form
results for one-point functions, multi-time correlators, antisymmetric functions,
and the OTOC--variance relation. Section~\ref{sec:heis} carries out the parallel
analysis for Heisenberg models and in Section~\ref{sec:IW} we discuss the In\"on\"u--Wigner contraction
that establishes the relation between the two cases via an algebra contraction.
Section~\ref{sec:HO} applies the formulas to the
harmonic oscillator with various initial states for the purposes of illustration.
Section~\ref{sec:cm} applies these general results to the complexity--momentum correspondence. 
Appendix~\ref{app:Heis-factor} discusses in detail a factorization formula in the $\mk h$ algebra.
Appendix~\ref{app:HO} collects detailed
expressions for the harmonic oscillator cases.

\section{Krylov spread complexity and Krylov correlators}
\la{sec:general}

Given the initial normalized quantum state $\ket{K_{0}}$ and the Hamiltonian operator $H$, we apply the Lanczos algorithm and define
the  states $\{\ket{K_{n}}\}_{n\ge 1}$ and Lanczos coefficients $\{a_{n}\}_{n\ge 0}$, $\{b_{n}\}_{n\ge 1}$ by iterating 
\bea
& \ket{A_{n+1}} = (H-a_{n})\ket{K_{n}}-b_{n}\ket{K_{n-1}}, \qquad \ket{K_{n}} = b_{n}^{-1}\ket{A_{n}},\\
& a_{n} =\mmm{K_{n}}{H}{K_{n}}, \qquad b_{n}=\braket{A_{n}}{A_{n}}^{1/2}.
\eea
The states $\{\ket{K_{n}}\}$ are orthonormal and, by construction, the action of $H$ is tridiagonal in their basis
(with convention $b_{0}=0$)
\be
\la{2.3}
H\ket{K_{n}} = a_{n}\ket{K_{n}}+b_{n}\ket{K_{n-1}}+b_{n+1}\ket{K_{n+1}}.
\ee
The survival amplitude is defined as 
\be
\la{2.4}
S(t) = \braket{K_{0}(t)}{K_{0}} = \mmm{K_{0}}{e^{iHt}}{K_{0}},
\ee
and is the generating function of moments
\be
\la{2.5}
\mu_{n}\equiv \mmm{K_{0}}{(iH)^{n}}{K_{0}} = \frac{d^{n}}{dt^{n}}S(t)|_{t=0}.
\ee
Using (\ref{2.3}), given the survival amplitude, the Lanczos coefficients are determined order by order by the moment relations
\footnote{The all-order solution is related to the Toda hierarchy \cite{Dymarsky:2019elm}, see also \cite{Dymarsky:2021bjq}.}
\be
\la{2.6}
\mu_{1} = ia_{0}, \qquad
\mu_{2} = -a_{0}^{2}-b_{1}^{2}, \qquad
\mu_{3} = -i [a_{0}^{3}+(2a_{0}+a_{1})b_{1}^{2}], \dots\ .
%\mu_{4} &= a_{0}^{4}+3a_{0}^{2}b_{1}^{2}+2a_{0}a_{1}b_{1}^{2}+(a_{1}^{2}+b_{1}^{2}+b_{2}^{2})b_{1}^{2},
\ee
%and so on. This gives
%\bea
%a_{0} &= -i\mu_{1}, \qquad
%b_{1} = \sqrt{\mu_{1}^{2}-\mu_{2}}, \\
%a_{1} &= i\frac{\mu_{1}^{3}-2\mu_{1}\mu_{2}+\mu_{3}}{\mu_{1}^{2}-\mu_{2}}, \qquad
%b_{2} = \frac{\sqrt{\mu_{2}^{3}+\mu_{3}^{2}+\mu_{1}^{2}\mu_{4}-\mu_{2}(2\mu_{1}\mu_{3}+\mu_{4})}}{\mu_{1}^{2}-\mu_{2}}, \quad\cdots.
%\eea

%The survival amplitude in (\ref{2.4}) may be written as the Fourier transform of a spectral measure
%\be
%S(t) = \int d\mu(E) e^{itE}, \qquad d\mu(E) = dE\, \sum_{n}\delta(E-E_{n})\, |\braket{E_{n}}{K_{0}}|^{2}.
%\ee
%One may introduce a family of orthogonal polynomials with respect to $d\mu(E)$
%\be
%\int d\mu(E)\, p_{n}(E)\, p_{m}(E) = \delta_{nm},\qquad \deg p_{n}=n,
%\ee
%and they obey a 3-term recursion with the Lanczos coefficients, see for instance \cite{Muck:2022xfc}, 
%\be
%Ep_{n} = a_{n}p_{n}+b_{n}p_{n-1}+b_{n+1}p_{n+1}.
%\ee

\paragraph{Krylov complexity and correlators}

The Krylov (spread) complexity of the state $\ket{K_{0}}$ is defined as 
\be
C(t) = \sum_{n=1}^{\infty}n\, |\braket{K_{n}}{K_{0}(t)}|^{2},
\ee
and is a one-point function of the Heisenberg operator $N(t) = e^{itH}Ne^{-itH}$, where $N\ket{K_{n}} = n\ket{K_{n}}$
is the Krylov number operator, \ie
\be
C(t) = \mmm{K_{0}}{N(t)}{K_{0}}.
\ee
Here, we consider the multi-point correlators 
\be
C_{p}(t_{1}, \dots, t_{p}) = \mmm{K_{0}}{N(t_{1})\cdots N(t_{p})}{K_{0}},
\ee
from which one can obtain any desired out-of-time-order correlator (OTOC)  in Krylov space.

\paragraph{Survival amplitude for the thermofield double state chain}

An important case, relevant in our holographic application, is when the initial state is 
the thermofield double state \cite{Maldacena:2001kr}. In general, it is 
the following entangled state in two copies (L, R) of the state space 
\be
\la{2.10}
\ket{\psi_{\beta}} = \frac{1}{\sqrt{Z_{\beta}}}\sum_{n}e^{-\frac{1}{2}\beta E_{n}}\ket{n}_{L}\otimes \ket{n}_{R}, \qquad Z_{\beta} = \sum_{n}e^{-\beta E_{n}}.
\ee
Here $\beta$ is the inverse temperature and tracing over either copy yields the thermal-density matrix with partition function $Z_{\beta}$. 
The state is annihilated by  $H_{L}-H_{R}$ and thus the spread complexity is the same whether one evolves with $H_{L}, H_{R}$, or $\frac{1}{2}(H_{L}+H_{R})$. Explicitly,
\be
\ket{\psi_{\beta}(t)} = e^{-iHt}\ket{\psi_{\beta}} = \frac{1}{\sqrt{Z_{\beta}}}\sum_{n}e^{-\frac{1}{2}(\beta+2it) E_{n}}\ket{n}_{L}\otimes \ket{n}_{R}.
\ee
The survival amplitude is then obtained from the analytic continuation of $\ket{\psi_{\beta}}$ and the partition function according to
\be
\la{2.12}
S(t) = \braket{\psi_{\beta+2it}}{\psi_{\beta}} = \frac{1}{Z_{\beta}}\sum_{n}e^{-\frac{1}{2}\beta E_{n}}e^{-\frac{1}{2}(\beta-2it) E_{n}} = \frac{Z_{\beta-it}}{Z_{\beta}}.
\ee

\section{Exact results in models with $\mk{sl}(2, \mathbb R)$ or Heisenberg symmetry}
\la{sec:math}

Before proceeding, we remark that despite the variety of exact results 
derived in this section, the underlying methods are straightforward: 
the algebraic structure of the Krylov chain reduces the computation 
of any multi-point correlator to matrix elements of Lie algebra 
generators, evaluated using standard representation theory. 
The resulting formulas are exact and capture non-trivial physical 
content, but the route to them is technically undemanding.

\subsection{The $\mk{sl}(2, \mathbb R)$ case}
\la{sec:sl2}

A class of quantum systems whose spread complexity is
analytically tractable was identified early on in \cite{Caputa:2021sib,
Balasubramanian:2022tpr}, see also Appendix A in \cite{Caputa:2024sux}.
Consider the algebra $\mk{sl}(2, \mathbb R)$ with generators $\{L_{-1}, L_{0}, L_{1}\}$ obeying
\be
[L_{n},L_{m}]=(n-m)L_{n+m}, \qquad n,m=-1,0,1.
\ee
Starting from the highest weight state $\ket{h}$ 
\be
L_{0}\ket{h} = h\ket{h}, \qquad L_{1}\ket{h}=0,
\ee
we can build  orthonormal states obtained by further application of $L_{-1}$  (notation is $\ket{h,0}\equiv\ket{h}$)
\be
\ket{h,n} = \sqrt\frac{\Gamma(2h)}{n!\Gamma(2h+n)}\,L_{-1}^{n}\ket{h}.
\ee
The action of the $\mk{sl}(2, \mathbb R)$ generators on these states reads
\bea
\la{3.4}
L_{0}\ket{h,n} &= (h+n)\ket{h,n}, \\
L_{1}\ket{h,n} &= \sqrt{n(n+2h-1)}\ket{h,n-1}, \qquad
L_{-1}\ket{h,n} = \sqrt{(n+1)(n+2h)}\ket{h,n+1}.
\eea
Thus, for a Hamiltonian operator of the form 
\be
\la{3.5}
H = \gamma L_{0}+\alpha(L_{1}+L_{-1})+\delta\, \mathbb{I},
\ee
we have the tridiagonal action ($\ket{h,-1}\equiv 0$)
\ba
& H\ket{h,n} = a_{n}\ket{h,n}+b_{n}\ket{h,n-1}+b_{n+1}\ket{h,n+1}, \\
\la{3.7}
& a_{n} = \gamma(h+n)+\delta, \qquad b_{n} = \alpha\sqrt{n(n+2h-1)}.
\ea
A key fact is that the states $\{\ket{h,n}\}_{n\ge 0}$ are precisely the Krylov chain $\{\ket{K_{n}}\}$ built 
by starting from the initial state $\ket{h}$.

\subsubsection{Spread complexity}

The standard calculation of $C(t)$ in $\mk{sl}(2,\mathbb R)$ models is as follows. We omit the $\delta$ term in (\ref{3.5}) since it does not enter complexity.
One starts with the following factorization of the temporal evolution operator, \footnote{Eq.~(\ref{3.8}) follows from the Lie algebra decoupling theorem, see for instance \cite{Chowdhury:2025tci}.}
\be
\la{3.8}
e^{-itH}=e^{-it(\gamma L_{0}+\alpha(L_{1}+L_{-1}))} = e^{\eta_{-1}L_{-1}}e^{\eta_{0}L_{0}}e^{\eta_{1}L_{1}},
\ee
where $\eta_{-1}, \eta_{0}, \eta_{1}$ are functions of $\gamma, \alpha$.
Using the explicit $2\times 2$ representation \footnote{This is enough to compute factorizations that depend only on the Lie algebra and not on the definition
of adjoint.  The fact that $L_{1}=-(L_{-1})^{\dag}$ is not relevant here.}
\be
L_{0}=\frac{1}{2}\begin{pmatrix}1 & 0 \\ 0 & -1 \end{pmatrix}, \qquad
L_{-1}=\begin{pmatrix}0 & 1 \\ 0 & 0 \end{pmatrix}, \qquad
L_{1}=\begin{pmatrix}0 & 0 \\ -1 & 0 \end{pmatrix},
\ee
we get 
\bea
\la{3.10}
\eta_{-1} &= \eta_{1} = -\frac{2\alpha}{\gamma-i D\coth(\frac{1}{2}Dt)} \qquad
\eta_{0} = -2\log\bigg[\cosh(\tfrac{1}{2}Dt)+\frac{i\gamma}{D}\sinh(\tfrac{1}{2}Dt)\bigg], \\
D &\equiv \sqrt{4\alpha^{2}-\gamma^{2}}.
\eea
This gives the following evolution of the initial state $\ket{K_{0}}$
\ba
\ket{K_{0}(t)} &= e^{-itH}\ket{h} = e^{\eta_{-1}L_{-1}}e^{\eta_{0}L_{0}}e^{\eta_{1}L_{1}}\ket{h}  %= e^{\eta_{0}h}\sum_{n=0}^{\infty}\frac{\eta_{-1}^{n}}{n!}L_{-1}^{n}\ket{h}\lp
= e^{\eta_{0}h}\sum_{n=0}^{\infty}\frac{\eta_{-1}^{n}}{n!}\sqrt\frac{n!\Gamma(2h+n)}{\Gamma(2h)}\ket{K_{n}}.
\ea
Thus, we obtain 
\be
|\braket{K_{n}}{K_{0}(t)}|^{2} = e^{(\eta_{0}+\eta_{0}^{*})h}|\eta_{-1}|^{2n}\frac{\Gamma(2h+n)}{n!\Gamma(2h)},
\ee
and the spread complexity is readily computed as 
\be
\la{3.13}
C(t) = \sum_{n=1}^{\infty}n|\braket{K_{n}}{K_{0}(t)}|^{2} = 2h\,e^{(\eta_{0}+\eta_{0}^{*})h}\,\frac{|\eta_{-1}|^{2}}{(1-|\eta_{-1}|^{2})^{2h+1}}.
\ee
Let us specialize to the case $\alpha, \gamma\in\mathbb R$, which ensures $H$ is hermitian.  This implies $D^{*}=\pm D$ depending on it being real or imaginary.
Using that $\eta_{-1},\eta_{0}, \eta_{1}$ are even in $D$, we get 
\ba
\eta_{0}+\eta_{0}^{*} &= -2\log\bigg[\cosh^{2}\frac{Dt}{2}+\frac{\gamma^{2}}{D^{2}}\sinh^{2}\frac{Dt}{2}\bigg], \qquad
|\eta_{-1}|^{2} = \frac{4\alpha^{2}}{\gamma^{2}+D^{2}\coth^{2}\frac{Dt}{2}}.
\ea
Substituting these expressions into (\ref{3.13}) gives the neat formula
\be
\la{3.15}
C(t) = h\,\frac{8\alpha^{2}}{4\alpha^{2}-\gamma^{2}}\sinh^{2}\frac{Dt}{2}.
\ee

\subsubsection{Krylov correlators}

We now rederive $C(t)$
using an algebraic approach that extends directly to multi-point correlators.
The spread complexity is 
\be
C(t) = \sum_{n=1}^{\infty}n|\braket{K_{n}}{K_{0}(t)}|^{2} = \sum_{n=1}^{\infty}\mmm{K_{0}}{e^{itH}}{K_{n}}\,n\,\mmm{K_{n}}{e^{-itH}}{K_{0}}.
\ee
If we introduce the Krylov number operator $N$ such that $N\ket{K_{n}} = n\ket{K_{n}}$, we can write the complexity as a one-point function 
\be
C(t) = \sum_{n=1}^{\infty}n|\braket{K_{n}}{K_{0}(t)}|^{2} = \mmm{K_{0}}{N(t)}{K_{0}}, \qquad N(t) = e^{itH}N e^{-itH}.
\ee
It is then natural to introduce multi-time $p$-point correlators 
%\footnote{
%An OTOC case could be instead 
%the expectation value of the commutator $\mmm{K_{0}}{[N(t_{1}), N(t_{2})]}{K_{0}}$ for $t_{2}>t_{1}$.
%Notice that this is zero for $t_{1}=0$. 
%}
\be
C_{p}(t_{1},\dots, t_{p}) =\mmm{K_{0}}{N(t_{1})\cdots N(t_{p})}{K_{0}}.
\ee
In order to explain how they can be computed, let us start by an alternative derivation of the spread complexity (\ref{3.15}).
Since $\ket{K_{n}}=\ket{h,n}\sim L^{n}_{-1}\ket{h}$,
we have 
\be
N\ket{h,n} = n\ket{h,n}\qquad \text{and thus}\qquad N = L_{0}-h.
\ee
For $H$ as in (\ref{3.5}), we need to evaluate
\be
L_{0}(t) = e^{itH}L_{0}e^{-itH}.
\ee
This can be done for all $L_{n}(t)$ using the equations of motion
\be
\dot L_{n}(t) = i\, [H, L_{n}(t)].
\ee
Their explicit form is 
\bea
\dot L_{0}(t) &=  i\alpha(L_{1}(t)-L_{-1}(t)), \\
\dot L_{1}(t) &= -i[\gamma L_{1}(t)+2\alpha L_{0}(t)], \\
\dot L_{-1}(t) &= i[\gamma L_{-1}(t)+2\alpha L_{0}(t)].
\eea
Solving these equations, we find in particular
\ba
\la{3.23}
L_{0}(t) &= -\frac{\gamma^{2}-4\alpha^{2}\cosh(Dt)}{D^{2}}\,L_{0}\lp
+\alpha\frac{2\gamma\sinh^{2}(\frac{1}{2}Dt)-iD\sinh(Dt)}{D^{2}}L_{-1}
+\alpha\frac{2\gamma\sinh^{2}(\frac{1}{2}Dt)+iD\sinh(Dt)}{D^{2}}L_{1}.
\ea
In the following, we will be interested also in correlators of the Krylov speed $N'(t) = L_{0}'(t)$ given by 
\ba
L_{0}'(t) &= \frac{4\alpha^{2}}{D}\sinh(Dt)\, L_{0}+\alpha\bigg(i\cosh(Dt)+\frac{\gamma}{D}\sinh(Dt)\bigg)\, L_{1}
+\alpha\bigg(-i\cosh(Dt)+\frac{\gamma}{D}\sinh(Dt)\bigg)\, L_{-1}.
\ea
The expression (\ref{3.23}) gives immediately the previous formula (\ref{3.15}),
\be
\la{3.25}
C(t) = \mmm{K_{0}}{[L_{0}(t)-h]}{K_{0}}  =\frac{8\alpha^{2}h}{D^{2}}\sinh^{2}(\tfrac{1}{2}Dt).
\ee
Let us move on to generalizations.

\paragraph{One-point functions $\langle N(t)^{p}\rangle$}

To compute the  higher order complexities
\be
C^{(p)}(t) = \sum_{n=0}^{\infty}n^{p}|\braket{K_{n}}{K_{0}(t)}|^{2} = \mmm{K_{0}}{[L_{0}(t)-h]^{p}}{K_{0}},
\ee
it is convenient to introduce a parameter $\mu$, write $L_{0}(t) = AL_{0}+(B+iC)L_{1}+(B-iC)L_{-1}$, where $A,B,C$ read off from (\ref{3.23}), and start from the factorization 
\be
e^{\mu(L_{0}(t)-h)} = e^{-\mu h}e^{\mu AL_{0}+\mu (B+iC)L_{1}+\mu (B-iC)L_{-1}} = e^{-\mu h}e^{F_{-1} L_{-1}}e^{F_{0}L_{0}}e^{F_{1}L_{1}},
\ee
where $F_{I}=F_{I}(t, \mu)$.
It follows that 
\be
C^{(p)}(t) = \bigg(\frac{d}{d\mu}\bigg)^{p}[e^{-\mu h+F_{0}(t, \mu)}]\bigg|_{\mu=0}.
\ee
From (\ref{3.10}) we have 
\be
\la{3.29}
F_{0}(t,\mu) = -2\log\bigg[\cosh\frac{\mu\sqrt{A^{2}-4(B^{2}+C^{2})}}{2}-\frac{A\sinh\frac{\mu\sqrt{A^{2}-4(B^{2}+C^{2})}}{2}}{\sqrt{A^{2}-4(B^{2}+C^{2})}}\bigg].
\ee
Using $A,B,C$ from (\ref{3.23}) gives
\be
F_{0}(t,\mu) = -2\log\bigg(\cosh\frac{\mu}{2}+\frac{\gamma^{2}-4\alpha^{2}\cosh(Dt)}{4\alpha^{2}-\gamma^{2}}\sinh\frac{\mu}{2}\bigg).
\ee
Now we use 
\ba
\exp\bigg\{-h\bigg[ & \mu+2\log\bigg(\cosh\frac{\mu}{2}+\rho\sinh\frac{\mu}{2}\bigg)\bigg]\bigg\} = \bigg(\frac{1}{1+\frac{1+\rho}{2}(e^{\mu}-1)}\bigg)^{2h}\lp
= \sum_{p=0}^{\infty}\frac{\mu^{p}}{p!}\sum_{k=0}^{p}\binom{2h+k-1}{k}k! S_{2}(p,k)\bigg(-\frac{1+\rho}{2}\bigg)^{k},
\ea
where $S_{2}(p,k)$ are Stirling numbers of the second kind. This gives the general formula
\be
\la{3.32}
C^{(p)}(t) =\sum_{k=1}^{p}\binom{2h+k-1}{k}k!\,  S_{2}(p,k)\bigg(\frac{4\alpha^{2}}{D^{2}}\sinh^{2}\frac{Dt}{2}\bigg)^{k},
\ee
where we used $S_{2}(p,0)=0$. One verifies that $p=1$ reproduces (\ref{3.15}). The  next cases are
\bea
\la{3.33}
C^{(2)}(t) &= \frac{8h\alpha^{2}}{D^{2}}\bigg[1+\frac{4(1+2h)\alpha^{2}}{D^{2}}\sinh^{2}\frac{Dt}{2}\bigg]\, \sinh^{2}\frac{Dt}{2}, \\
C^{(3)}(t) &= \frac{8h\alpha^{2}}{D^{2}}\bigg[
1+\frac{12(1+2h)\alpha^{2}}{D^{2}}\sinh^{2}\frac{Dt}{2}+\frac{32(1+h)(1+2h)\alpha^{4}}{D^{4}}\sinh^{4}\frac{Dt}{2}
\bigg]\,\sinh^{2}\frac{Dt}{2},
\eea
and so on.

\paragraph{One-point functions $\langle N'(t)^{p}\rangle$}

In the same way we can compute the one-point function $\langle [N'(t)]^{p}\rangle$, \ie
\be
\wt C^{(p)}(t) =  \mmm{K_{0}}{[L_{0}'(t)]^{p}}{K_{0}}.
\ee
We use again (\ref{3.29}) with $A\to A'$, $B\to B'$, $C\to C'$, where $A',B',C'$
are read off from the expression for $L_0'(t)$ above (\ref{3.25}). This gives
\be
F_{0}(t,\mu) = -2\log\bigg[\cos(\alpha\mu)-\frac{2\alpha}{D}\sinh(Dt)\sin(\alpha\mu)\bigg].
\ee
Now we use the identity 
\be
\la{3.36}
\frac{1}{(\cos x+\rho \sin x)^{2h}} = \sum_{n=0}^{\infty}\frac{x^{n}}{n!}(-2ih)^{n}\sum_{j=0}^{n}\frac{1}{h^{j}}\binom{n}{j}\sum_{\ell=0}^{j}S_{2}(j,\ell)(2h)_{\ell}\bigg(-\frac{1+i \rho}{2}\bigg)^{\ell},
\ee
where $(2h)_{\ell} = \Gamma(2h+\ell)/\Gamma(2h)$ is the Pochhammer symbol. This gives the Taylor expansion of $e^{F_{0}}$ and thus
%\separator
%\ba
%\exp\bigg\{-2h & \log[\cos(\alpha\mu)+\rho\sin(\alpha\mu)]\bigg\} = \frac{1}{(\cos(\alpha\mu)+\rho\sin(\alpha\mu))^{2h}}\lp
%= e^{2ih\alpha \mu}\bigg[1+\frac{1-i\rho}{2}(e^{2i\alpha\mu}-1)\bigg]^{-2h}.
%\ea
%We have the expansion 
%\bs
%\be
%\frac{e^{hx}}{(1+\xi(e^{x}-1))^{2h}} = \sum_{n=0}^{\infty}(-1)^{n}\frac{x^{n}}{n!}\sum_{j=0}^{n}\binom{n}{j}h^{n-j}\sum_{\ell=0}^{j}S_{2}(j,\ell)(-1)^{\ell}\frac{\Gamma(2h+\ell)}{\Gamma(2h)}
%(1-\xi)^{\ell}.
%\ee
%\es
%Hence
%\ba
%\exp\bigg\{-2h & \log[\cos(\alpha\mu)+\rho\sin(\alpha\mu)]\bigg\} \lp
%=  \sum_{n=0}^{\infty}(-1)^{n}\frac{(2i\alpha\mu)^{n}}{n!}\sum_{j=0}^{n}\binom{n}{j}h^{n-j}\sum_{\ell=0}^{j}S_{2}(j,\ell)(-1)^{\ell}\frac{\Gamma(2h+\ell)}{\Gamma(2h)}
%\bigg(\frac{1+i\rho}{2}\bigg)^{\ell},
%\ea
the explicit formula \footnote{This is a generalization of the $h=1/2$ case studied in \cite{boyadzhiev2007derivative}. }
\be
\la{3.37}
\wt C^{(p)}(t) =  (-2i\alpha)^{p}\sum_{j=0}^{p}\binom{p}{j}h^{p-j}\sum_{\ell=0}^{j}S_{2}(j,\ell)(-1)^{\ell}\frac{\Gamma(2h+\ell)}{\Gamma(2h)}
\bigg(\frac{1-i\frac{2\alpha}{D}\sinh(Dt)}{2}\bigg)^{\ell}.
\ee
The  first cases are (the first is consistent with $\wt C(t) = C'(t)$)
\bea
\wt C(t) &= \frac{4h\alpha^{2}}{D}\sinh(Dt), \\
\wt C^{(2)}(t) &= 2h\alpha^{2}\bigg[1+\frac{4(1+2h)\alpha^{2}}{D^{2}}\sinh^{2}(Dt)\bigg], \\
\wt C^{(3)}(t) &= \frac{8h\alpha^{4}}{D}\bigg[1+3h+\frac{4(1+h)(1+2h)\alpha^{2}}{D^{2}}\sinh^{2}(Dt)\bigg]\sinh(Dt).
\eea
Notice that 
Eq.~(\ref{3.37}) is an explicit representation of the Taylor expansion of the l.h.s. It is manifestly real despite 
the convenient use of complex numbers. A simple manifest real expansion -- but not explicit -- 
may be found by noticing that in (\ref{3.36}) the polynomials $P_{n}(\rho)$ defined by 
\be
\sum_{n=0}^{\infty}P_{n}\frac{x^{n}}{n!} = (\cos x+\rho \sin x)^{-2h},
\ee 
obeys the recursion 
\be
P_{n+1} = -(1+\rho^{2}) P'_{n}-2h\rho P_{n}, \qquad P_{0}=1,
\ee 
and thus we have the (implicit) manifestly real expression
\be
P_{n}=(-1)^{n}(1+\rho^{2})^{-h}\bigg[(1+\rho^{2})\frac{d}{d\rho}\bigg]^{n}(1+\rho^{2})^{h}.
\ee
Alternatively, using $\rho=-\tan t$, this reads simply $P_{n} = \cos^{2h}t(\frac{d}{dt})^{n}\sec^{2h}t$.

\paragraph{Two-point function and antisymmetrized higher correlators}

The 2-point function with generic $t_{1}$, $t_{2}$ is 
\ba
\mmm{K_{0}}{N(t_{1})N(t_{2})}{K_{0}}.
\ea
Using (\ref{3.23}) and the action of generators on the Krylov basis in  (\ref{3.4}) we get \footnote{
Notice that $C_{2}(t_{1}, t_{2})\sim h^{2}$ for large $h$, but this is canceled in the connected correlator
$C_{2}(t_{1}, t_{2})-C(t_{1})\,C(t_{2}) \sim h$.
}
\ba
C_{2}(t_{1}, t_{2}) &= \frac{8h\alpha^{2}}{D^{4}}\sinh\frac{Dt_{1}}{2}\sinh\frac{Dt_{2}}{2}\lp
\bigg[(D^{2}-2(1+2h)\alpha^{2})\cosh\frac{D(t_{1}-t_{2})}{2}
+2(1+2h)\alpha^{2}\cosh\frac{D(t_{1}+t_{2})}{2}-i\, D\gamma\,\sinh\frac{D(t_{1}-t_{2})}{2}\bigg].
\ea
In the same way, one can consider more insertions of $N(t)$. 
Another set of quantities that we are going to discuss in some detail are the \textit{fully antisymmetrized} correlators
\be
\la{3.44}
A_{p}(t_{1}, \dots, t_{p}) = \sum_{\sigma\in S_{p}}(-1)^{\sigma}\mmm{K_{0}}{N(t_{\sigma_{1}})\cdots N(t_{\sigma_{p}})}{K_{0}} .
\ee
Remarkably, they are non-zero only for $p=1,2,3$, the case $p=1$ being trivial. 
Indeed writing $L_{0}(t) = f_{0}(t) L_{0}+f_{1}(t)L_{1}+f_{-1}(t)L_{-1}$, the antisymmetrized sum is a product of 
$p$ factors from $\{f_{0},f_{-1},f_{1}\}$ at $p$ times $t_{1}, \dots, t_{p}$. By antisymmetrization we get zero if $p\ge 4$. The remaining non-trivial cases are then  $p=2,3$
and we get for them the expressions
\bea
\la{3.45}
A_{2}(t_{1}, t_{2}) &= -\frac{16i h\alpha^{2}\gamma}{D^{3}}\sinh\frac{Dt_{1}}{2}\sinh\frac{Dt_{12}}{2}\sinh\frac{Dt_{2}}{2},\\ 
A_{3}(t_{1}, t_{2}, t_{3}) &=  \frac{16i h \alpha^{2}\gamma}{D^{3}}\sinh\frac{Dt_{12}}{2}\sinh\frac{Dt_{13}}{2}\sinh\frac{Dt_{23}}{2},
\eea
where $t_{ij}\equiv t_{i}-t_{j}$.
In the following, it will be interesting to differentiate in time to get antisymmetrized correlators of $N'(t)$. This gives
\ba
\la{3.46}
\wt{A}_{2}(t_{1}, t_{2})  &= \partial_{t_{1}}\partial_{t_{2}}A_{2}(t_{1}, t_{2}) = -\frac{4i h\alpha^{2}\gamma}{D}\sinh\frac{Dt_{12}}{2},\\ 
\la{3.47}
\wt{A}_{3}(t_{1}, t_{2},t_{3})  &= \partial_{t_{1}}\partial_{t_{2}}\partial_{t_{3}}A_{3}(t_{1}, t_{2}, t_{3}) = 0.
\ea
Comparing with (\ref{3.25}),  we note the relation 
\be
i\,\partial_{t_{1}}\partial_{t_{2}}A_{2}(t_{1}, t_{2}) = \gamma\, C'(t_{12}).
\ee
The  vanishing in (\ref{3.47}) is not accidental and is related to the $\mk{sl}(2,\mathbb R)$ structure. 
Indeed, one may consider for a generic smooth $f(t)$ the functional equation 
\be
\partial_{t_{1}}\partial_{t_{2}}\partial_{t_{3}} [f(t_{12})f(t_{13})f(t_{23})] = 0.
\ee
Setting $x=t_{12}$ and $t_{2}=t_{3}$ we get 
\be
f'(x)[f''(x)f(0)-f(x)f''(0)]=0.
\ee
A trivial solution is $f(x)$ constant. The other solutions have the form $f(x) = c_{1}e^{\kappa\,x}+c_{2}e^{-\kappa\,x}$,
with constant $c_{1},c_{2}, \kappa$.
Substituting into the initial equation, we find that any $c_{1},c_{2}, \kappa$ work (and this includes the case of constant $f$).
Our case has indeed $f(x) = A \sinh(B x)$ from $\mk{sl}(2, \mathbb R)$ algebra and it is of the special form leading to vanishing.

\subsubsection{Relation with quartic OTOC}

The quartic OTOC, first introduced in a condensed matter context 
\cite{larkin1969quasiclassical} and later identified as a sharp 
diagnostic of quantum chaos and holographic scrambling 
\cite{Maldacena:2015waa}, is 
in this context the following expectation value of a squared commutator:
\be
\text{OTOC}(t) = \mmm{K_{0}}{[N(t), N(0)]^{2}}{K_{0}}.
\ee
As a consequence of $\mk{sl}(2,\mathbb R)$  algebra, we may prove the simple relation
\be
\la{3.52}
\text{OTOC}(t) = -\text{Var}(t), 
\ee
where we introduced the notation 
\be
\text{Var}(t) \equiv C^{(2)}(t)-C(t)^{2}.
\ee
To prove (\ref{3.52}), we first observe that from $N(0)\ket{K_{0}}=0$ (in general $N(0)\ket{K_{n}}=n\ket{K_{n}}$ by definition) we have
\be
\text{OTOC}(t)  = -\mmm{K_{0}}{N(t)N(0)^{2}N(t)}{K_{0}} = -\Vert N(0)N(t)\ket{K_{0}}\Vert^{2}.
\ee
Now, we use the $\mk{sl}(2,\mathbb R)$ representation structure to write
\be
N(t)\ket{K_{0}} = c_{0}(t)\, \ket{K_{0}}+c_{1}(t)\, \ket{K_{1}},
\ee
and in particular
\be
N(0)N(t)\ket{K_{0}} = c_{1}(t)\, \ket{K_{1}}.
\ee
This gives
\be
\text{OTOC}(t)  = -|c_{1}(t)|^{2}.
\ee
On the other hand, we also have
\be
\text{Var}(t) = \Vert N(t)\ket{K_{0}}\Vert^{2}-\mmm{K_{0}}{N(t)}{K_{0}}^{2} = |c_{0}(t)|^{2}+|c_{1}(t)|^{2}-|c_{0}(t)|^{2}=|c_{1}(t)|^{2},
\ee
and (\ref{3.52}) follows.
%
%\separator
%To prove (\ref{3.52}), we use $N(0) = L_{0}-h$ and adopt the general parametrization 
%$N(t) = -h+(1+t f_{0}(t))L_{0}+t f_{1}(t)L_{1}+t f_{-1}(t) L_{-1}$ that makes the $t=0$ limit manifest. 
%From $N(0)\ket{K_{0}} =  0$ we can compute 
%\be
%\text{OTOC}(t)  = -\mmm{K_{0}}{N(t)N(0)^{2}N(t)}{K_{0}} = -2h t^{2}f_{1}(t)f_{-1}(t).
%\ee
%On the other hand
%\ba
%C(t) &= t f_{0}(t), \\
%C^{(2)}(t) &= \mmm{K_{0}}{(-h+(1+t f_{0}(t))L_{0}+t f_{1}(t)L_{1}+t f_{-1}(t) L_{-1})^{2}}{K_{0}} \lp
%= t^{2}f_{0}(t)^{2}+t^{2}f_{1}(t)f_{-1}(t)\mmm{K_{0}}{L_{1}L_{-1}+L_{-1}L_{1}}{K_{0}} \lp
%= C(t) ^{2}+2ht^{2}f_{1}(t)f_{-1}(t),
%\ea
%and (\ref{3.52}) follows. 
Notice that the explicit general  expression for the variance is 
\be
\text{Var}(t) =  \frac{8h\alpha^{2}}{D^{2}}\bigg(1+\frac{4\alpha^{2}}{D^{2}}\sinh^{2}\frac{Dt}{2}\bigg)\, \sinh^{2}\frac{Dt}{2}.
\ee
We can introduce similar quantities involving $N'$, \ie 
\ba
\wt{\text{OTOC}}(t) = \mmm{K_{0}}{[N'(t), N'(0)]^{2}}{K_{0}}, \qquad \wt{\text{Var}}(t) \equiv \wt C^{(2)}(t)-\wt C(t)^{2},
\ea
In this case $N'(0)\ket{K_{0}}\neq 0$ --- unlike $N(0)\ket{K_{0}}=0$ which was 
the key input in the proof of (\ref{3.52}) --- and the squared commutator 
$[N'(t),N'(0)]^{2}$ does not reduce to a simple norm. 
The evaluation of $\wt{\text{OTOC}}$ requires the full four-term expansion 
of the squared commutator and knowledge of $N'(t)$ on the entire Krylov chain, 
so it is more direct to proceed by explicit computation. We compute
\ba
\wt{\text{OTOC}}(t)  &= 16h\frac{\alpha^{4}(2\alpha^{2}+h \gamma^{2})}{D^{2}}\sinh^{2}(Dt), \qquad
\wt{\text{Var}}(t)  = 2h\alpha^{2}\bigg(1+\frac{4\alpha^{2}}{D^{2}}\sinh^{2}(Dt)\bigg).
\ea
and thus verify the exact relation 
\be
\wt{\text{OTOC}}(t)+2(2\alpha^{2}+h\gamma^{2})(-2h\alpha^{2}+\wt{\text{Var}}(t) ) = 0.
\ee
Unlike (\ref{3.52}), this relation is not universal but depends explicitly 
on the model parameters $\alpha$ and $\gamma$, reflecting the fact that 
no parameter-free algebraic shortcut analogous to the $N$ case is available.

\subsection{The Heisenberg case}
\la{sec:heis}

The $\mk{sl}(2,\mathbb R)$ models are closely related to models based on the quantum oscillator algebra. 
Let us consider the Heisenberg algebra $\mk h$ \footnote{This is more precisely the oscillator algebra:
the Heisenberg-Weyl algebra generated by $\{a, a^{\dag}, \mathbb{I}\}$, augmented by the number operator.}
generated by $a, a^{\dag}$ and $\hat n = a^{\dag}a$ with 
\be
[a, a^{\dag}]=1, \qquad [\hat n, a] = -a, \qquad [\hat n, a^{\dag}] = a^{\dag}.
\ee
The occupation number states are 
\be
\ket{n} = \frac{(a^{\dag})^{n}}{\sqrt{n!}}\ket{0}, \qquad a\ket{n}=\sqrt{n}\,\ket{n-1}, \qquad a^{\dag}\ket{n} = \sqrt{n+1}\,\ket{n+1}.
\ee
The following Hamiltonian built with the algebra generators
\be
\la{3.64}
H = \omega\, a^{\dag}a+\l(a+a^{\dag})+\delta,
\ee
gives a tridiagonal action with coefficients
\be
\la{3.65}
a_{n}=\omega n+\delta, \qquad b_{n}=\l\sqrt{n},
\ee
and starting from $\ket{0}$, the Krylov chain is precisely $\ket{K_{n}} = \ket{n}$. The time evolved Krylov number is 
\be
N(t) = e^{itH}a^{\dag}a e^{-itH} = a^{\dag}(t) a(t).
\ee
From the equations of motion 
\ba
\dot a(t) = i[H,a] = i\,(-\l-\omega a),
\ea
we obtain 
\ba
\la{3.68}
a(t) &= e^{-it\omega}a-\l\frac{1-e^{-it\omega}}{\omega}, \qquad 
a^{\dag}(t) = e^{it\omega}a^{\dag}-\l\frac{1-e^{it\omega}}{\omega}. 
\ea

\subsubsection{Krylov correlators}
\paragraph{One-point functions $\langle N(t)^{p}\rangle$}
The 1-point function admits a more direct derivation 
than in the $\mk{sl}(2,\mathbb R)$ case. From (\ref{3.68}), 
\ba
\la{3.69}
C(t) &= \bra{0}\bigg(e^{it\omega}a^{\dag}-\l\frac{1-e^{it\omega}}{\omega}\bigg)\bigg(e^{-it\omega}a-\l\frac{1-e^{-it\omega}}{\omega}\bigg)\ket{0}
= \frac{4\l^{2}}{\omega^{2}}\sin^{2}\frac{\omega t}{2}.
\ea
The higher complexities may be evaluated similarly, but
it is convenient to study their generating function. To this end, one may start from the factorization
\be
e^{\mu \hat n+A\,a+\bar A\,a^{\dag}} = e^{f}e^{\bar g a^{\dag}}e^{s\hat n}e^{g a},
\ee
where we need to find $f,g,\bar g, s$ in terms of $\mu$ and $A$. Once these functions are known, the generating function is
\be
\la{3.71}
\mmm{0}{e^{\mu \hat n+A\,a+\bar A\,a^{\dag}}}{0} = e^{f}.
\ee
We collect in Appendix \ref{app:Heis-factor} this calculation as well as a 
simpler direct derivation of the specific matrix element (\ref{3.71}). The result is 
\be
\la{3.72}
\mmm{0}{e^{\mu \hat n+A\,a+\bar A\,a^{\dag}}}{0} = \exp\bigg[|A|^{2}\frac{e^{\mu}-1-\mu}{\mu^{2}}\bigg].
\ee
It follows that 
\be
\mmm{0}{e^{\mu N(t)}}{0} = \exp\bigg[\frac{4\l^{2}}{\omega^{2}}\sin^{2}\frac{\omega t}{2}(e^{\mu}-1)\bigg],
\ee
and this implies the general formula -- rather simpler than  (\ref{3.32}) --
\be
\la{3.74}
C^{(p)}(t) =\sum_{k=1}^{p}S_{2}(p,k)\bigg(\frac{4\l^{2}}{\omega^{2}}\sin^{2}\frac{\omega t}{2}\bigg)^{k}.
\ee
The first cases are 
\bea
\la{3.75}
C^{(2)}(t) &= \frac{4\l^{2}}{\omega^{4}}\bigg(\omega^{2}+4\l^{2}\sin^{2}\frac{\omega t}{2}\bigg)\sin^{2}\frac{\omega t}{2}, \\
C^{(3)}(t) &= \frac{4\l^{2}}{\omega^{6}}\bigg(\omega^{4}+12\l^{2}\omega^{2}\sin^{2}\frac{\omega t}{2}+16\l^{4}\sin^{4}\frac{\omega t}{2}\bigg)\sin^{2}\frac{\omega t}{2}.
\eea

\paragraph{One-point functions $\langle N'(t)^{p}\rangle$}

For the matrix elements of $[N'(t)]^{p}$ the determination of the generating function is simpler because
\be
N'(t) = i[H,\hat n(t)] = i[\l(a(t)+a^{\dag}(t))+\omega \hat n(t)+\delta, \hat n(t)] = i\l(a(t)-a^{\dag}(t)),
\ee
so that 
\be
e^{\mu N'} = e^{i\mu \l[a(t)-a^{\dag}(t)]} = \exp\bigg(i\l\mu e^{-it\omega}a-i\l\mu e^{it\omega}a^\dag\bigg)\exp\bigg(\frac{2\l^{2}\mu}{\omega}\sin\omega t\bigg).
\ee
Using
\be
\exp(-i\l\mu e^{it\omega}a^\dag)\exp(i\l\mu e^{-it\omega}a) = \exp(i\l\mu e^{-it\omega}a-i\l\mu e^{it\omega}a^\dag)\exp\bigg(-\frac{1}{2}\l^{2}\mu^{2}\bigg),
\ee
we get 
\be
e^{\mu N'} = \exp\bigg[\frac{2\l^{2}\mu}{\omega}\sin\omega t+\frac{1}{2}\l^{2}\mu^{2}\bigg]e^{(\cdots) a^{\dag}}e^{(\cdots)a},
\ee
where the terms in round brackets will not be needed. We obtain 
\ba
\la{3.80}
\wt C^{(p)}(t) &= (\frac{d}{d\mu})^{p}\exp\bigg[\frac{2\l^{2}\mu}{\omega}\sin\omega t+\frac{1}{2}\l^{2}\mu^{2}\bigg]\bigg|_{\mu=0}
= (\frac{i\l}{\sqrt 2})^{p}H_{p}\bigg(-i\l\sqrt{2}\frac{ \sin(\omega t)}{\omega}\bigg),
\ea
where $H_{p}$ are Hermite polynomials. The first cases are 
\bea
\wt C(t) &= \frac{2\l^{2}}{\omega}\sin(\omega t), \qquad
\wt C^{(2)}(t) = \l^{2}\bigg(1+\frac{4\l^{2}}{\omega^{2}}\sin^{2}(\omega t)\bigg),\\
\wt C^{(3)}(t) &= \frac{2\l^{4}}{\omega}\bigg(3+\frac{4\l^{2}}{\omega^{2}}\sin^{2}(\omega t)\bigg)\sin(\omega t).
\eea

\paragraph{Higher-point functions}

Higher-point functions can be computed similarly, and one gets for instance 
\ba
C_{2}(t_{1},t_{2}) &= \bra{0}\bigg(e^{it_{1}\omega}a^{\dag}-\l\frac{1-e^{it_{1}\omega}}{\omega}\bigg)\bigg(e^{-it_{1}\omega}a-\l\frac{1-e^{-it_{1}\omega}}{\omega}\bigg)\lp
\bigg(e^{it_{2}\omega}a^{\dag}-\l\frac{1-e^{it_{2}\omega}}{\omega}\bigg)\bigg(e^{-it_{2}\omega}a-\l\frac{1-e^{-it_{2}\omega}}{\omega}\bigg)\ket{0}\lp
= \frac{4\l^{2}}{\omega^{4}}\bigg[(2\l^{2}+\omega^{2})\cos\frac{\omega(t_{1}-t_{2})}{2}-2\l^{2}\cos\frac{\omega(t_{1}+t_{2})}{2}\bigg]\,
\sin\frac{\omega t_{1}}{2}\sin\frac{\omega t_{2}}{2}\lp
-4i\frac{\l^{2}}{\omega^{2}}\sin\frac{\omega t_{1}}{2}\sin\frac{\omega(t_{1}-t_{2})}{2}\sin\frac{\omega t_{2}}{2}.
\ea
In particular, the antisymmetrized 2-point function is 
\be
\la{3.83}
A_{2}(t_{1},t_{2}) = C_{2}(t_{1},t_{2})-C_{2}(t_{2},t_{1}) = -8i\frac{\l^{2}}{\omega^{2}}\sin\frac{\omega t_{1}}{2}\sin\frac{\omega(t_{1}-t_{2})}{2}\sin\frac{\omega t_{2}}{2}.
\ee
The antisymmetrized 3-point function, computed analogously, is
\be
\la{3.84}
A_{3}(t_{1},t_{2},t_{3}) =8i\frac{\l^{2}}{\omega^{2}}\sin\frac{\omega t_{12}}{2}\sin\frac{\omega t_{13}}{2}\sin\frac{\omega t_{23}}{2}.
\ee
Higher-point antisymmetrized functions vanish, $A_{p}=0$ for $p\ge 4$.

\subsection{Large $h$ limit and In\"on\"u-Wigner  contraction $\mk{sl}(2,\mathbb R)\to \mk h$}
\la{sec:IW}

The results for the $\mk{h}$ models can alternatively be obtained by a suitable algebraic contraction of 
an associated model with $\mk{sl}(2, \mathbb R)$ algebra, providing a consistency check.
The action of $\mk{sl}(2,\mathbb R)$ generators in (\ref{3.4}) can be written 
\bea
(L_{0}-h)\ket{h,n} &= n\ket{h,n}, \\
\frac{1}{\sqrt{2h}}L_{1}\ket{h,n} &= \sqrt{n}\,\bigg(1+\frac{n-1}{2h}\bigg)^{1/2}\ket{h,n-1}, \\
\frac{1}{\sqrt{2h}}L_{-1}\ket{h,n} &= \sqrt{(n+1)}\bigg(1+\frac{n}{2h}\bigg)^{1/2}\ket{h,n+1},
\eea
and we have the algebra
\be
[L_{0}-h, \frac{1}{\sqrt{2h}}L_{\pm 1}] = \mp \frac{1}{\sqrt{2h}}L_{\pm 1}, \qquad [\frac{1}{\sqrt{2h}}L_{1},\frac{1}{\sqrt{2h}}L_{-1}] = \frac{1}{h}(L_{0}-h)+1.
\ee
This means that we may identify for $h\gg 1$ 
\be
L_{0}-h \to \hat n, \qquad \frac{1}{\sqrt{2h}}L_{1} \to a, \qquad  \frac{1}{\sqrt{2h}}L_{-1} \to a^{\dag}.
\ee
The $\mk{sl}(2,\mathbb R)$ Hamiltonian is 
\be
H = \gamma L_{0}+\alpha(L_{1}+L_{-1}) = \gamma\hat n+\alpha\sqrt{2h}(a+a^{\dag})+\gamma h.
\ee
Comparing with (\ref{3.64}), this implies the following identification valid at large $h$ -- constant terms are irrelevant for our purposes since
they do not affect the Krylov complexity or correlators -- 
\be
\gamma = \omega, \qquad \alpha = \frac{\l}{\sqrt{2h}}.
\ee
In particular, the combination $D$ in (\ref{3.10}) is  transformed into 
\be
D = \sqrt{4\alpha^{2}-\gamma^{2}} \to i\omega.
\ee
As an example of the map between complexities, we can consider, \cf (\ref{3.25}), 
\be
C(t) =\frac{8\alpha^{2}h}{4\alpha^{2}-\gamma^{2}}\sinh^{2}(\tfrac{1}{2}t\sqrt{4\alpha^{2}-\gamma^{2}}) \to \frac{4\l^{2}}{\omega^{2}}\sin^{2}\frac{\omega t}{2},
\ee
which agrees with (\ref{3.69}). Similarly one can check that (\ref{3.74}) follows from (\ref{3.32}) and that (\ref{3.80})  follows from (\ref{3.37}).
Notice that this also implies the validity of the OTOC-Variance relation (\ref{3.52}) in models with Heisenberg algebra.

\subsection{Krylov correlators in the harmonic oscillator with various initial states}
\la{sec:HO}

As an illustration of the above general formulas, we can consider the quantum harmonic oscillator with different choices for the initial state. For a unit mass oscillator
with angular frequency $\omega$, we consider three cases
\begin{enumerate}
\item Thermofield double state at inverse temperature $\beta$,
\item Coherent state with parameter $z$,
\item Squeezed state or initial gaussian with generic width $\sim \exp(-x^{2}/(4r))$.
\end{enumerate}
As reviewed in Appendix \ref{app:HO}, cases (1) and (3) have algebra $\mk{sl}(2, \mathbb R)$ while case (2) is described by a Heisenberg model.
The relevant parameters of the Hamiltonians (\ref{3.5}) and (\ref{3.64}) are summarized as follows
\be
\la{3.92}
\def\arraystretch{1.3}
\begin{array}{ccl}
\toprule
\textsc{Initial state} & \textsc{Algebra} & \textsc{Parameters}  \\
\midrule
\text{TFD} & \mk{sl}(2,\mathbb R) &   h = \frac{1}{2}, \qquad \gamma = \frac{\omega}{\tanh(\beta\omega/2)}, \qquad \alpha = \frac{\omega}{2\sinh(\beta\omega/2)}\\
\text{Coherent} & \mk{h} &  \l=\omega\,|z| \\
\text{Gaussian} & \mk{sl}(2,\mathbb R) & h = \frac{1}{4}, \qquad \gamma = \frac{1+4r^{2}\omega^{2}}{2r}, \qquad \ \ \alpha = \frac{1-4r^{2}\omega^{2}}{4r}
  \\
\bottomrule
\end{array}
\ee
The three cases exhibit several noteworthy features, highlighted below; 
detailed expressions are collected in Appendix~\ref{app:HO}.

\medskip\noindent
\textit{TFD initial state.}
With $h=1/2$ and $4\alpha^{2}-\gamma^{2}=-\omega^{2}$ (so $D=i\omega$, purely imaginary), 
the system lies in the oscillatory regime. 
The spread complexity and all higher one-point functions are purely oscillatory,
\be
C(t) = \frac{\sin^{2}\frac{\omega t}{2}}{\sinh^{2}\frac{\omega\beta}{2}},
\ee
with amplitude controlled by the thermal factor $\sinh^{-2}(\omega\beta/2)$.
The higher complexities $C^{(p)}(t)$ are polynomials in $C(t)$ itself, a direct
consequence of the Stirling-number structure of (\ref{3.32}), so all moments are
determined by the single oscillating combination $\sin^{2}(\omega t/2)/\sinh^{2}(\omega\beta/2)$.
The antisymmetrized functions $A_{2}$ and $A_{3}$ are non-zero and proportional to 
$\gamma = \omega/\tanh(\beta\omega/2)$, consistent with the factor of $\gamma$ 
 in (\ref{3.45}).

\medskip\noindent
\textit{Coherent initial state.}
The coherent state with amplitude $|z|$ is the unique case governed by the 
Heisenberg algebra with $\l=\omega|z|$, corresponding to Lanczos coefficients
$b_{n}=\l\sqrt{n}$ and diagonal entries $a_{n}=\omega n + \delta$ (with $\delta$ 
an irrelevant constant shift that does not enter complexity).
All Krylov correlators are purely oscillatory with frequency $\omega$ and are
controlled by the single amplitude $4|z|^{2}\sin^{2}(\omega t/2)$.
The higher complexities $C^{(p)}(t)$ are polynomials in $C(t)$ with coefficients 
growing with $p$, and their leading behavior at large $|z|$ reflects the 
Poissonian spread of the coherent state over the Krylov basis.
The antisymmetric functions $A_{2}$ and $A_{3}$ are non-zero and take the universal 
factorized form in (\ref{3.83}) and (\ref{3.84}) with amplitude $\sim |z|^{2}$; 
unlike the TFD case there is no thermal suppression, and the amplitude is 
determined purely by the coherent-state displacement $|z|$.

\medskip\noindent
\textit{Squeezed (Gaussian) initial state.}
The squeezed state returns to the $\mk{sl}(2,\mathbb R)$ framework with the 
 weight $h=1/4$,
corresponding to the metaplectic (oscillator) representation generated by 
$a^{2}$, $a^{\dag 2}$, and $a^{\dag}a$.
A notable feature is that the Krylov chain wave functions involve only even 
Hermite polynomials $H_{2n}$, reflecting a parity selection rule: the Hamiltonian 
preserves the parity sector of the initial state, effectively halving the chain.
The complexity and all correlators vanish identically when $r=1/(2\omega)$, \ie when 
the squeezed state coincides with the oscillator ground state, since 
$\alpha=(1-4r^{2}\omega^{2})/(4r)=0$ at that point and both $C^{(p)}$ and the 
antisymmetric amplitudes are proportional to $(1-4r^{2}\omega^{2})^{2}$.

\section{Krylov correlators and complexity--momentum correspondence}
\la{sec:cm}

We now examine what information can be gained by studying the exact Krylov correlators
in the context of holographic complexity--momentum correspondence, as outlined in the Introduction.
The specific setup we adopt is the same as in  \cite{Caputa:2024sux}, \ie the eternal two-sided AdS black hole.
Let us briefly recall the main facts. At the classical level, an eternal black hole has two disconnected exterior regions joined by an
Einstein--Rosen bridge --- a non-traversable wormhole.
Each exterior region has  its own asymptotic AdS boundary, and by the AdS/CFT dictionary
each boundary hosts a copy of the dual CFT.
Maldacena \cite{Maldacena:2001kr} showed that the quantum state corresponding
to this geometry is the thermofield double (TFD): a specific entangled pure state
of the two boundary CFTs that reproduces a thermal density matrix upon tracing out either
copy, \cf (\ref{2.10}). 
Thus, entanglement between the two decoupled boundary CFTs 
is geometrically realised in the bulk as the Einstein--Rosen 
bridge connecting the two exterior regions.

\subsection{The BTZ/CFT$_{2}$ dual pair}

A realization of this system is the BTZ black hole in AdS$_3$ \cite{Banados:1992wn,Banados:1992gq}. It is a quotient of
 AdS$_3$ whose holographic dual is precisely the TFD of two copies of 2d conformal theories,
 each living on a spatial circle. The Hawking temperature of the bulk black hole matches the temperature
$T = 1/\beta$ of the boundary thermal state.
%, and the Bekenstein--Hawking entropy
%equals the entanglement entropy between the two CFT's in the TFD state. 
The central charge of the two CFT's is $c=3L_{\rm AdS}/(2G_{N})$
where $L_{\rm AdS}$ is the AdS$_{3}$ radius.

On the boundary, the time-evolved initial Krylov state is obtained by exciting the thermofield double by a local primary 
operator $\mc O$  with dimension $\Delta=h+\bar h$ inserted at spatial position $x_{0}$ \cite{Berenstein:2019tcs}, 
and evolving with $H$
\be
\la{4.1}
\ket{K_{0}(t)} = \mc N e^{-itH}e^{-\UV H}\mc O_{\Delta}(x_{0})\ket{\text{TFD}_{\beta}},
\ee
where $\mc N$ is a normalization constant.
The parameter $\UV$ is a UV regulator. From now on, for brevity, we drop the subscript
and  write $\eps\equiv \UV$ in what follows. The conserved energy of the state $\ket{K_{0}(t)}$ is   
$E=\int dx \mmm{K_{0}}{T_{00}(x)}{K_{0}} = \Delta/\eps$. 
The Krylov coefficients for the chain associated with (\ref{4.1}) are obtained by using (\ref{2.12}), computing moments, and 
solving the moment relations (\ref{2.6}).
This is possible because the survival amplitude in (\ref{2.4}) may be computed by conformal symmetry in terms of the 2-point function ratio \cite{Caputa:2023vyr}
\footnote{The complex conjugate appears due to our convention for $S(t)$ in (\ref{2.4}).}
\bea
\la{4.2}
& S(t)^{*} = \frac{\langle\mc O(z_{1},\bar z_{1})\mc O(z_{2}(t), \bar z_{2}(t))\rangle}{\langle\mc O(z_{1},\bar z_{1})\mc O(z_{2}(0),\bar z_{2}(0))\rangle}, \\
&z_{1} = x_{0}+i\eps, \quad \bar z_{1} = x_{0}-i\eps, \quad z_{2}(t) = x_{0}-i(\eps+it), \quad \bar z_{2}(t) = x_{0}+i(\eps+it),
\eea
with the finite-temperature two-point function on the cylinder 
\be
\langle \mc O(z_{1}, \bar z_{1})\mc O(z_{2}, \bar z_{2})\rangle = \bigg[\frac{\beta}{2\pi}\sinh\frac{\pi z_{12}}{\beta}\bigg]^{-2h}
\bigg[\frac{\beta}{2\pi}\sinh\frac{\pi \bar z_{12}}{\beta}\bigg]^{-2\bar h}.
\ee
After substituting the explicit coordinates in (\ref{4.2}), this gives the  expression
\be
S(t) = \bigg(\frac{\sinh\frac{\pi(t+2i\eps)}{\beta}}{\sinh\frac{2\pi i \eps}{\beta}}\bigg)^{-2\Delta}.
\ee
As shown in \cite{Caputa:2024sux}, the corresponding  Lanczos coefficients read
\be
a_{n} = \frac{2\pi}{\beta\tan\frac{2\pi \eps}{\beta}}(n+\Delta), \qquad b_{n}=\frac{\pi}{\beta\sin\frac{2\pi \eps}{\beta}}\sqrt{n(n+2\Delta-1)}.
\ee
%If we use the 2-point function for a CFT of finite length $L$
%\be
%\langle \mc O(z_{1}, \bar z_{1})\mc O(z_{2}, \bar z_{2})\rangle = \bigg[\frac{L}{\pi}\sin\frac{\pi z_{12}}{L}\bigg]^{-2h}
%\bigg[\frac{L}{\pi}\sin\frac{\pi \bar z_{12}}{L}\bigg]^{-2\bar h},
%\ee
%we obtain 
%\be
%S(t) = \bigg(\frac{\sin\frac{\pi(t+2i\eps)}{L}}{\sin\frac{2\pi i \eps}{L}}\bigg)^{-2\Delta},
%\ee
%with Lanczos coefficients
%\be
%a_{n} = \frac{2\pi}{L\tanh\frac{2\pi \eps}{L}}(n+\Delta), \qquad b_{n}=\frac{\pi}{L\sinh\frac{2\pi \eps}{L}}\sqrt{n(n+2\Delta-1)}.
%\ee
Comparing with (\ref{3.7}), we see that the Krylov chain reproduces the $\mk{sl}(2,\mathbb R)$ representation structure with 
parameters
\ba
\la{4.6}
& h=\Delta, \qquad \alpha = \frac{\pi}{\beta\sin\frac{2\pi \eps}{\beta}}, \qquad \gamma= \frac{2\pi}{\beta\tan\frac{2\pi \eps}{\beta}}.
%\qquad D =\frac{2\pi}{\beta} ,
%\text{finite } L: \qquad & h=\Delta, \qquad \alpha = \frac{\pi}{L\sinh\frac{2\pi \eps}{L}}, \qquad \gamma= \frac{2\pi}{L\tanh\frac{2\pi \eps}{L}}, 
%\qquad D = \frac{2\pi i}{L}.
\ea
Notice that the  $D$ parameter is then real
\be
D =\sqrt{4\alpha^{2}-\gamma^{2}} = \frac{2\pi}{\beta},
\ee
and corresponds to an exponential growth in time of the spread complexity.

\subsection{Krylov correlation functions}

The Krylov complexity is given by the general formula (\ref{3.25}) that reads in our case
\ba
C(t) = 2\Delta\, \frac{\sinh^{2}\frac{\pi t}{\beta}}{\sin^{2}\frac{2\pi \eps}{\beta}}. 
\ea
Higher Krylov moments follow from the general formula (\ref{3.32})
\be
\la{4.9}
C^{(p)}(t) =\sum_{k=0}^{p}\binom{2\Delta+k-1}{k}k!\,  S_{2}(p,k)\bigg(\frac{\sinh^{2}\frac{\pi t}{\beta}}{\sin^{2}\frac{2\pi \eps}{\beta}}\bigg)^{k}.
\ee
The first cases are 
\ba
C^{(2)}(t) &= 2\Delta\bigg[1+(1+2\Delta)\frac{\sinh^{2}\frac{\pi t}{\beta}}{\sin^{2}\frac{2\pi\eps}{\beta}}\bigg]\, \frac{\sinh^{2}\frac{\pi t}{\beta}}{\sin^{2}\frac{2\pi\eps}{\beta}}, \\
C^{(3)}(t) &= 2\Delta\bigg[1+3(1+2\Delta)\frac{\sinh^{2}\frac{\pi t}{\beta}}{\sin^{2}\frac{2\pi\eps}{\beta}}+2(1+\Delta)(1+2\Delta)\frac{\sinh^{4}\frac{\pi t}{\beta}}{\sin^{4}\frac{2\pi\eps}{\beta}}\bigg]\, 
\frac{\sinh^{2}\frac{\pi t}{\beta}}{\sin^{2}\frac{2\pi\eps}{\beta}}.
\ea
For the one-point functions of $N'(t)$ we use (\ref{3.37}) with the parameters
(\ref{4.6}). The first  cases are 
\ba
\la{4.12}
\wt C(t) &= C'(t) = \Delta\frac{2\pi}{\beta}\frac{\sinh\frac{2\pi t}{\beta}}{\sin^{2}\frac{2\pi \eps}{\beta}}, \\
\wt C^{(2)}(t) &= \Delta\frac{2\pi^{2}}{\beta^{2}}\bigg[1+(1+2\Delta)\frac{\sinh^{2}\frac{2\pi t}{\beta}}{\sin^{2}\frac{2\pi \eps}{\beta}}\bigg]\,\frac{1}{\sin^{2}\frac{2\pi \eps}{\beta}}, \\
\wt C^{(3)}(t) &= \Delta\frac{4\pi^{3}}{\beta^{3}}\bigg[1+3\Delta+(1+\Delta)(1+2\Delta)\frac{\sinh^{2}\frac{2\pi t}{\beta}}{\sin^{2}\frac{2\pi \eps}{\beta}}\bigg]\,\frac{\sinh\frac{2\pi t}{\beta}}{\sin^{2}\frac{2\pi \eps}{\beta}}.
\ea
The antisymmetrized 2-point function is 
\be
A_{2}(t_{1},t_{2}) = -4i\Delta\frac{\cos\frac{2\pi \eps}{\beta}}{\sin^{3}\frac{2\pi \eps}{\beta}}\sinh\frac{\pi t_{1}}{\beta}\sinh\frac{\pi t_{12}}{\beta}\sinh\frac{\pi t_{2}}{\beta},
\ee
while the  antisymmetrized 3-point function reads
\be
A_{3}(t_{1},t_{2},t_{3}) = 4i\Delta\frac{\cos\frac{2\pi \eps}{\beta}}{\sin^{3}\frac{2\pi \eps}{\beta}}\sinh\frac{\pi t_{12}}{\beta}\sinh\frac{\pi t_{13}}{\beta}\sinh\frac{\pi t_{23}}{\beta}.
\ee
%
%
%\separator
%
%For higher one-point functions, the leading term at small $\eps$ and fixed time is 
%\bea
%\la{8.12}
%C^{(2)}(t) &= \frac{\beta^{4}}{4\pi^{4}\eps^{4}}\Delta(\Delta+\tfrac{1}{2})\, \sinh^{4}\frac{\pi t}{\beta}+\dots,\\
%C^{(3)}(t) &= \frac{\beta^{6}}{8\pi^{6}\eps^{6}}\Delta(\Delta+\tfrac{1}{2})(\Delta+1)\, \sinh^{6}\frac{\pi t}{\beta}+\dots, \\
%C^{(4)}(t) &= \frac{\beta^{8}}{16\pi^{8}\eps^{8}}\Delta(\Delta+\tfrac{1}{2})(\Delta+1)(\Delta+\tfrac{3}{2})\, \sinh^{8}\frac{\pi t}{\beta}+\dots, 
%\eea
%and so on. Using (\ref{3.32}) gives
%\be
%\la{4.9}
%C^{(p)}(t) =\sum_{k=0}^{p}\binom{2\Delta+k-1}{k}k!\,  S_{2}(p,k)\bigg(\frac{\sinh^{2}\frac{\pi t}{\beta}}{\sin^{2}\frac{2\pi \eps}{\beta}}\bigg)^{k},
%\ee
%and the leading term for $\eps\to 0$ is the term $k=p$ that gives the general formula
%\ba
%C^{(p)}(t) &=
%\bigg(\frac{\beta}{2\pi \eps}\bigg)^{2p}\binom{2\Delta+p-1}{p}p!\,  S_{2}(p,p)\sinh^{2p}\frac{\pi t}{\beta}+\cdots\lp 
%=\bigg(\frac{\beta}{2\pi \eps}\bigg)^{2p}\frac{\Gamma(2\Delta+p)}{\Gamma(2\Delta)}\sinh^{2p}\frac{\pi t}{\beta}+\cdots,
%\ea
%in agreement with (\ref{8.12}). 
%

\subsection{Probing the correspondence with Krylov correlators}

Our aim is to examine the Krylov correlators computed in the previous section in the semiclassical limit $\Delta\gg 1$
that allows a bulk interpretation for the complexity $C(t)$.

The insertion of the local operator $e^{-\eps H}\mc O(x_{0})$ at $t = 0$ is dual to a localised point particle in the bulk with mass $m\simeq \Delta$ for $\Delta\gg 1$ and $\eps \ll 1$,
located at $x=x_{0}$ and $z=\eps$ (in Poincar\'e coordinates) and at rest \cite{Nozaki:2013wia,Berenstein:2019tcs}. Consistently,
for large $\Delta$  the relative fluctuations of the energy of the state (\ref{4.1}) are suppressed as $\sim \Delta^{-1/2}$, at fixed UV cutoff,
confirming the semiclassical nature of the state.

The dual geometry of the thermal double state with inverse temperature $\beta$ is the AdS$_{3}$ black hole with redefined 
radial coordinate \footnote{The precise embedding may be found in \cite{Caputa:2024sux}. Notice that $\rho$ measures the proper (geodesic) distance from the horizon. }
\be
ds^{2} = g_{\mu\nu}dX^{\mu}dX^{\nu} = d\rho^{2}+\frac{4\pi^{2}}{\beta^{2}}(-\sinh^{2}\rho\, dt^{2}+\cosh^{2}\rho\, dx^{2}).
\ee
The equation of motion for a particle with mass $m$ is obtained from the variation of the static gauge action
\be
S =  \int dt \mc L, \qquad \mc L = -m \sqrt{-g_{\mu\nu}(X)\dot X^{\mu}\dot X^{\nu}}.
\ee
The radial motion is found by solving 
\be
\rho''-2\coth\rho\, (\rho')^{2}+\frac{2\pi^{2}}{\beta^{2}}\sinh(2\rho)=0,\qquad\Rightarrow\qquad
\tanh\rho(t) = c_{1}\sech(\frac{2\pi t}{\beta}+c_{2}).
\ee
Setting $c_{2}=0$, the particle is initially at rest. The other constant is fixed by imposing that the particle's initial position depends
on the $\eps$ regulator according to $\rho(0) = \log(\beta/(\pi \eps))$. This gives
\be
c_{1} = \tanh\log\frac{\beta}{\pi\eps} = \frac{1-\frac{\pi^{2}}{\beta^{2}}\eps^{2}}{1+\frac{\pi^{2}}{\beta^{2}}\eps^{2}}.
\ee
The proper momentum of the falling particle and its conserved energy are (the negative sign reflecting the inward falling)
\be
\la{4.21}
P=\frac{\partial \mc L}{\partial \rho'} = -\frac{m\beta}{2\pi \eps}\sinh\frac{2\pi t}{\beta}+O(\eps), \qquad
E = \rho'\frac{\partial \mc L}{\partial \rho'}-\mc L = \frac{m}{\eps}+O(\eps).
\ee
We identify $m=\Delta$ in the semiclassical 
regime $\Delta\gg 1$.

Denoting for brevity $\langle \mc O \rangle \equiv \mmm{K_{0}}{\mc O}{K_{0}}$, from (\ref{4.12}) we read
\be
\langle N'(t)\rangle = \Delta\frac{2\pi}{\beta}\frac{\sinh\frac{2\pi t}{\beta}}{\sin^{2}\frac{2\pi \eps}{\beta}} = \frac{\beta\Delta}{2\pi\eps^{2}}\sinh\frac{2\pi t}{\beta}+\cdots,
\ee
where omitted terms are subleading at small $\eps$ and will be neglected. Using (\ref{4.21}) this gives the known result, \cf (\ref{1.2}),
\be
\la{4.23}
\langle N'(t)\rangle = -\frac{1}{\eps}\, P.
\ee
We now turn to the Krylov moments $C^{(p)} = \langle N(t)^{p}\rangle$.  
Using (\ref{4.21}), we  express $t$ in terms of proper momentum, 
\be
\la{4.24}
t = -\frac{\beta}{2\pi}\text{arcsinh}\, \frac{2\pi \eps P}{\beta\Delta} = -\frac{\eps}{\Delta}\, P + \cdots,
\ee
and  analyze (\ref{4.9}) by expanding first in $\eps\to 0$ and then in large $\Delta$. Using the relations 
\ba
\frac{\sinh^{2}\frac{\pi t}{\beta}}{\sin^{2}\frac{2\pi \eps}{\beta}} &= \frac{P^{2}}{4\Delta^{2}}+O(\eps),\qquad
\binom{2\Delta+k-1}{k} = \frac{(2\Delta)^{k}}{k!}+\cdots
\ea
we get 
\be
\la{4.26}
C^{(p)} =\sum_{k=1}^{p}\frac{(2\Delta)^{k}}{k!}k!\,  S_{2}(p,k)\bigg(\frac{P}{2\Delta}\bigg)^{2k} = \frac{P^{2}}{2\Delta}+O(1/\Delta^{2}).
\ee
Note that the leading term $P^{2}/(2\Delta)$ is independent of $p$. All higher moments collapse to the same function of $P$ at leading order,
so the semiclassical limit is blind to $p$.
A similar calculation gives
\be
\la{4.27}
\frac{d}{dt}C^{(p)} = -\frac{P}{\eps}+O(1/\Delta),
\ee
where only the subleading terms at large $\Delta$ depend on $p$. This clearly follows from (\ref{4.26}), using the first term in the expansion (\ref{4.24}). 
In more detail, we may use
\be
\frac{d}{dt}\frac{\sinh^{2}\frac{\pi t}{\beta}}{\sin^{2}\frac{2\pi \eps}{\beta}} = -\frac{P}{2\Delta \eps}+O(\eps),
\ee
and compute 
\ba
\frac{d}{dt}C^{(p)}(t) &=\sum_{k=0}^{p}\binom{2\Delta+k-1}{k}k!\,  S_{2}(p,k)\, k\, \bigg(\frac{\sinh^{2}\frac{\pi t}{\beta}}{\sin^{2}\frac{2\pi \eps}{\beta}}\bigg)^{k-1}\frac{d}{dt}\frac{\sinh^{2}\frac{\pi t}{\beta}}{\sin^{2}\frac{2\pi \eps}{\beta}}\lp
= \sum_{k=0}^{p}\frac{(2\Delta)^{k}}{k!}k!\,  S_{2}(p,k)\, k\, \bigg(\frac{P^{2}}{4\Delta^{2}}\bigg)^{k-1}\bigg(-\frac{P}{2\Delta\eps}\bigg).
\ea
The leading term at large $\Delta$ is the $k=1$ contribution that gives (\ref{4.27}). The case $p=1$ recovers the previous result  (\ref{4.23}).

The above calculation shows that in the 
semiclassical limit $\Delta\gg 1$, $\varepsilon\to 0$, all higher complexities 
$C^{(p)}$ carry identical leading bulk information, collapsing to the same 
function $P^{2}/(2\Delta)$ of the proper radial momentum regardless of $p$. This is a first hint that 
it is non-trivial to establish a dictionary for Krylov correlators. 
Still, the variance of $N'(t)$ provides information consistent 
with the semiclassical particle picture.
One has the general expression in terms of $\mk{sl}(2,\mathbb R)$ parameters
\be
\wt{\text{Var}}(t) \equiv \langle[N'(t)]^{2}\rangle-\langle N'(t)\rangle^{2}= 2h\alpha^{2}\bigg(1+\frac{4\alpha^{2}}{4\alpha^{2}-\gamma^{2}}\sinh^{2}(Dt)\bigg).
\ee
In our case, it reads
\be
\wt{\text{Var}}(t)  = \Delta\frac{2\pi^{2}}{\beta^{2}}\bigg[1+\frac{\sinh^{2}\frac{2\pi t}{\beta}}{\sin^{2}\frac{2\pi \eps}{\beta}}\bigg]\frac{1}{\sin^{2}\frac{2\pi \eps}{\beta}}.
\ee
Taking the bulk limit $\eps\to 0$, $\Delta\gg 1$, and expressing time in terms of momentum,  this reduces to 
\be
\la{4.32}
\wt{\text{Var}}(t)  = \frac{\Delta}{2\eps^{2}}+\cdots = \frac{E^{2}}{2\Delta}+\cdots,
\ee
where we recall that $E=\Delta/\eps$.
At fixed energy, this is suppressed at large $\Delta$ implying that fluctuations of the Krylov speed correspond in the bulk 
to quantities that are beyond the leading semiclassical regime. 
This is consistent with the particle picture: the semiclassical state has a narrow energy distribution,
which translates into narrow fluctuations of the Krylov speed.

The two-point function provides a  more interesting case: the 
operator nature of $N$ and $N'$ at different times forbids a direct geometric bulk interpretation.
The out-of-time ordered correlator of Krylov speed
\be
\wt C_{2}(t_{1},t_{2}) = \langle N'(t_{1}) N'(t_{2}) \rangle, 
\ee
has no symmetry under $t_{1}\leftrightarrow t_{2}$, an exchange that corresponds to  Hermitian conjugation.
Its general expression in terms of $\mk{sl}(2,\mathbb R)$ parameters is:
\ba
\wt C_{2}(t_{1}, t_{2}) &= 2h\alpha^{2}\bigg(1-\frac{2(1+2h)\alpha^{2}}{D^{2}}\bigg)\cosh(D(t_{1}-t_{2}))+\frac{4h(1+2h)\alpha^{4}}{D^{2}}\cosh(D(t_{1}+t_{2}))\lp
-\frac{2ih\alpha^{2}\gamma}{D}\sinh(D(t_{1}-t_{2})).
\ea
Its bulk limit can be computed and reads
\be
\la{4.35}
\langle N'(t_{1}) N'(t_{2}) \rangle = \frac{1}{\eps^{2}}\bigg[\frac{\Delta}{2}+P_{1}P_{2}+\frac{i}{2}(P_{1}-P_{2})+O(1/\Delta)\bigg]+O(\eps^{0}).
\ee
It follows that the antisymmetric part is simple and independent on $\Delta$ at leading order (we write only the most singular term for $\eps\to 0$)
\be
\la{4.36}
\langle [N'(t_{1}), N'(t_{2})] \rangle = i\, \frac{1}{\eps^{2}}(P_{1}-P_{2})+O(1/\Delta).
\ee
Thus,  the commutator of two Krylov speeds has a well defined semiclassical limit 
and corresponds to the proper momentum difference of the falling particle  at the two times where $N'(t)$ are computed. In this case the two sides of (\ref{4.36})
are antisymmetric and the momentum difference is associated with the non-commutativity of $N'(t)$ at different times. 
The r.h.s. in (\ref{4.36}) is imaginary, consistent with the fact that the 
Hermitian conjugate of the commutator in the l.h.s. is the same as exchanging $t_{1}, t_{2}$ and thus $P_{1}, P_{2}$. 
Notice that the r.h.s. is linear in momentum while
the l.h.s. is quadratic in $N'$. Again, it is not possible to establish a simple dictionary relating $N'(t)$ to $P$ and the duality 
probes the non-commutativity of $N'(t)$ at different times,
which is not captured by any classical single-particle observable.
Still, (\ref{4.36}) establishes a semiclassical correspondence for the 2-point antisymmetric function of the Krylov speed, 
extending (\ref{1.2}) to a two-time observable. 

Naively, the 3-point (fully) antisymmetric function of $N'(t)$ could be a candidate for an antisymmetric combination of $P_{1}, P_{2}, P_{3}$. However, 
our previous results imply its vanishing due to the underlying $\mk{sl}(2, \mathbb R)$ structure
\be
\la{4.37}
\partial_{t_{1}}\partial_{t_{2}}\partial_{t_{3}}A_{3}(t_{1},t_{2},t_{3})=0,
\ee
and it is thus not expressible as a non-trivial combination of the three associated proper momenta. A formula similar to (\ref{4.36})
is obtained by considering the general  three-point function 
\ba
\langle & N'(t_{1})\, N'(t_{2})\, N'(t_{3})\rangle = 
-\frac{8 h (1+2 h) \alpha ^4 (2 (-1+h) \alpha ^2+\gamma ^2)}{D^{3}} \cosh (D(t_{1}-t_{3}))\sinh (Dt_{2})\lp
+\frac{8 h \alpha ^4 (2 (1+h+2 h^2) \alpha ^2+h \gamma ^2) }{D^{3}}\cosh (D(t_{1}+t_{3})) \sinh (Dt_{2} )\lp
-\frac{8 i h (1+2 h) \alpha ^4 \gamma}{D^{2}}  \sinh (Dt_{2}) \sinh (D(t_{1}-t_{3}))
+\frac{8 h^2 \alpha ^4}{D} \cosh (Dt_{2}) \sinh (D(t_{1}+t_{3})).
\ea
Substituting the parameters (\ref{4.6}) and taking the bulk limit $\eps\to 0$, $\Delta \gg 1$, one obtains
 the following relation for the partially antisymmetrized correlator 
\be
\la{4.39}
\langle N'(t_{1})\,N'(t_{2})\,N'(t_{3})-N'(t_{3})\,N'(t_{2})\, N'(t_{1})\rangle = -2i\, \frac{1}{\eps^{3}}\, P_{2}\, (P_{1}-P_{3})+O(1/\Delta),
\ee
where the r.h.s. has a well defined non-trivial limit for $\Delta\gg 1$ and has the right antisymmetry under $1\leftrightarrow 3$ as in the l.h.s.
The construction in (\ref{4.39}) can be generalized to higher-point functions. As an example, 
starting from the general four-point function $\langle  N'(t_{1})\, N'(t_{2})\, N'(t_{3})\, N'(t_{4})\rangle$ --- that we can compute by the 
methods used so far, but is rather cumbersome --- one gets 
\be
\la{4.40}
\langle [N'(t_{1}), N'(t_{2})]\, [N'(t_{3}), N'(t_{4})] \rangle = -\frac{1}{\eps^{4}}\, (P_{1}-P_{2})\, (P_{3}-P_{4})+O(1/\Delta),
\ee
where we notice that this expression is real, consistent with Hermitian conjugation corresponding  to the combined exchange $1\leftrightarrow 3$ and $2\leftrightarrow 4$
that leaves the r.h.s. invariant.

We note that our exact formulas provide explicit $1/\Delta$ corrections 
to the leading results (\ref{1.7})--(\ref{1.9}). For instance, 
from (\ref{4.35}), the next-order correction to (\ref{1.7}) is
\be
i\,\langle [N'(t_{1}), N'(t_{2})] \rangle = 
-\frac{P_{1}-P_{2}}{\eps^{2}}+\frac{1}{\Delta}\,\frac{2\pi^{2}}{\beta^{2}}\,P_{1}P_{2}(P_{1}-P_{2})+O(1/\Delta^{2}).
\ee
These are exact CFT predictions that could in principle be matched 
to quantum corrections to the probe particle's trajectory in the 
BTZ background, computed from the worldline quantum mechanics 
of the dual bulk field. Such a matching would provide a non-trivial 
test of the correspondence beyond the semiclassical level, 
within the regime of classical bulk geometry.

We close with a cautionary remark. The clean bulk interpretation found in 
(\ref{4.36}), (\ref{4.39}), and (\ref{4.40}) relies on several features that 
are special to the present setup: the exact $\mk{sl}(2,\mathbb R)$ symmetry of 
the Krylov chain, the semiclassical limit $\Delta\gg 1$, and the specific 
partial symmetrizations of the multi-time correlators that were identified 
precisely because they admit a simple bulk expression. It is not clear 
whether analogous results hold in higher dimensions, for finite $\Delta$, or 
for CFTs without an underlying $\mk{sl}(2,\mathbb R)$ structure. The 
correspondences found here may therefore be specific to the AdS$_3$/CFT$_2$ 
setting and should be regarded as indicative rather than as evidence for a 
universal dictionary.

\section{Conclusions}

In this paper we have studied multi-time Krylov correlators --- natural extensions of spread complexity ---
in quantum systems with $\mk{sl}(2, \mathbb R)$ or Heisenberg-Weyl symmetry, and probed their holographic
interpretation.
These are analogous to out-of-time-order correlators, but built directly in Krylov space, 
thereby inheriting the simplicity of the  Krylov chain construction.

We considered model Hamiltonians with emergent $\mk{sl}(2,\mathbb R)$ symmetry
(or Heisenberg, related by algebra contraction) and derived exact closed-form 
results for the Krylov correlators. For the equal-time moments $C^{(p)}(t)$ 
we obtained a general formula valid for all $p$, expressed in terms of 
Stirling numbers of the second kind. For the multi-time correlators 
$C_p(t_1,\ldots,t_p)$ we derived explicit results for the two- and 
three-point functions and their antisymmetrized combinations. 
A structural result is the exact relation between the quartic OTOC 
$\langle[N(t),N(0)]^2\rangle$ and the variance of the Krylov number, 
$\text{OTOC}(t) = -\text{Var}(t)$. 
The antisymmetrized 3-point function factorizes into a product of 
three $\sinh$ functions of the time differences, and higher-point 
antisymmetric functions vanish as a consequence of the 
$\mk{sl}(2,\mathbb R)$ structure.

A direct application of our results is to the holographic 
duality between the BTZ black hole in AdS$_{3}$ and 
two non-interacting 2d conformal field theories.
In this context, the dual boundary state is the thermofield double state at the
Hawking temperature. At the semiclassical level, a point particle falling in the 
gravitational field of the black hole  is dual to the excitation of the thermofield double
by a local primary operator whose conformal dimension equals  the particle mass. 
The Krylov complexity of this state grows exponentially in time and its time derivative
is proportional to the proper momentum of the falling particle.
Applying our general results, we find that in general it is not possible to establish a direct dictionary relating the Krylov speed operator
$N'(t)$ to radial proper momentum. 
However, we identified specific correlators that have a simple 
semiclassical bulk interpretation of that kind. In particular, 
the Krylov antisymmetric 2-point function is proportional to the proper momentum difference between the two 
times of the insertions, $\langle [N'(t_{1}), N'(t_{2})]\rangle\, \propto\, P_{1}-P_{2}$.  
Similarly, in the case of the 3-point function, we presented a specific partially 
antisymmetrized cubic correlator of $N'$ at three times, expressible in terms of three proper momenta.
This suggests a continuing pattern, and indeed we presented an analogous result for a quartic correlator of $N'$ at four times built as 
the product of two commutators $\langle [N'(t_{1}), N'(t_{2})][N'(t_{3}), N'(t_{4})]\rangle\,\propto\, (P_{1}-P_{2})(P_{3}-P_{4})$.

These results develop the boundary side of the correspondence, which has 
received comparatively little attention since \cite{Caputa:2024sux} despite 
extensive work on the bulk side. Still, the correspondences found in (\ref{1.7})--(\ref{1.9}) rely on the exact 
$\mk{sl}(2,\mathbb R)$ symmetry of the Krylov chain and the semiclassical 
limit $\Delta\gg 1$, and should be regarded as indicative of a broader 
dictionary rather than evidence for a universal one.

A deeper extension of the correspondence, going beyond the semiclassical limit, would proceed in two steps.
At the level of the classical bulk geometry, the natural candidates 
are quantum corrections to the probe dynamics at order $1/\Delta$, 
whose boundary counterparts are already provided by our exact formulas. 
Going beyond the classical geometry altogether, off-shell contributions 
to the gravitational path integral --- such as replica wormholes --- 
may encode the $p$-dependence of the higher Krylov moments $C^{(p)}$, 
as suggested in \cite{Fu:2025kkh}.

In the present AdS$_{3}$/CFT$_{2}$ setting quantizing the bulk geometry is 
technically demanding, but in the context of the JT gravity/double-scaled SYK duality the 
situation may be more tractable: there the chord-number basis provides a discrete 
bulk counterpart to the Krylov chain \cite{Rabinovici:2023yex,Ambrosini:2024sre}, 
and the gravitational path integral is under sufficient control that off-shell 
contributions may in principle be accessible. 
Steps in this direction have recently been taken in 
\cite{Heller:2024ldz,Fu:2025kkh}: the former shows that 
the Krylov/C=V match extends to the full quantum regime 
in sine dilaton gravity, \footnote{
See also \cite{Aguilar-Gutierrez:2026ogo,Aguilar-Gutierrez:2026nmd} for a recent analysis of complexity in DSSYK deformed models. 
}
while the latter argues that 
$C^{(p)}$ for $p\ge 2$ captures replica wormhole contributions.
In this context, the radial proper momentum / rate of growth of Krylov complexity correspondence has been studied in the DSSYK model with matter in 
\cite{Aguilar-Gutierrez:2025pqp}.
%
%
%
%A step in this direction within the JT gravity framework has 
%recently been taken in \cite{Fu:2025kkh}, where it is argued that $C^{(p)}$ 
%for $p\ge 2$ captures connected bulk contributions encoded by replica wormholes 
%in the JT gravity dual, while a logarithmic variant of Krylov complexity 
%probes the replica saddle structure. 
We regard this as the most 
promising direction for future work on the holographic interpretation of 
higher Krylov correlators beyond the semiclassical level.
The results of this paper provide the exact boundary correlators against which any such bulk 
construction must ultimately be matched. 

A more ambitious complementary direction would be to study spread complexity 
in a fully-fledged AdS$_3$/CFT$_2$ string theory embedding, 
such as type IIB on AdS$_3\times S^3\times T^4$ dual to the 
D1-D5 system \cite{Maldacena:1997re}. The D1-D5 system 
contains operators --- such as twist operators of the symmetric 
orbifold --- whose two-point functions encode genuine dynamical 
information beyond universal conformal kinematics. Both Krylov 
complexity for such operators and its possible relation to 
string corrections in the bulk remain an open problem.

\section*{Acknowledgements}

We thank P. Nandy, S. E. Aguilar-Gutierrez,  A. Grabarits, and J. Xu for useful comments.
MB is supported by the INFN grant GAST. EA is
supported by the MUR project GINEVRA, prot. 2022BZYBWM.

\appendix

\section{Factorization in the $\mk h$ algebra}
\la{app:Heis-factor}

We derive here the factorization formula used in Section~\ref{sec:heis} to compute generating functions of Krylov correlators in the Heisenberg case.
It reads
\be
\la{A.1}
e^{\mu \hat n+A\,a+\bar A\,a^{\dag}} = e^{f}e^{\bar g a^{\dag}}e^{s\hat n}e^{g a},
\ee
where the ordering on the right is chosen to simplify the vacuum matrix element. In (\ref{A.1}) we want
to determine $f,g,\bar g, s$ as functions of $\mu$ and $A$. 
To get them, we introduce a parameter $\tau$ and compute 
\be
\la{A.2}
\partial_{\tau}e^{\tau(\mu \hat n+A\,a+\bar A\,a^{\dag})} = (\mu \hat n+A\,a+\bar A\,a^{\dag})\, e^{\tau(\mu \hat n+A\,a+\bar A\,a^{\dag})}.
\ee
The r.h.s. with $f=f(\tau)$ etc. is 
\ba
\la{A.3}
f' e^{f}e^{\bar g a^{\dag}}& e^{s\hat n}e^{g a}+e^{f}\bar g' a^{\dag} e^{\bar g a^{\dag}}e^{s\hat n}e^{g a}
+e^{f}e^{\bar g a^{\dag}}s' \hat n e^{s\hat n}e^{g a}+e^{f}e^{\bar g a^{\dag}}e^{s\hat n}g' a e^{g a} \lp
= \bigg[f'+\bar g'\, a^{\dag}+s'\, e^{\bar g a^{\dag}}\hat n e^{-\bar g a^{\dag}}
+g'\, e^{\bar g a^{\dag}}e^{s \hat n}\, a\, e^{-s \hat n}e^{-\bar g a^{\dag}} \bigg]e^{\tau(\mu \hat n+A\,a+\bar A\,a^{\dag})}.
\ea
Now we use the relations -- they can be proved by differentiating in $\bar  g$ or $s$ -- 
\ba
& e^{\bar g a^{\dag}}\hat n e^{-\bar g a^{\dag}} = \hat n -\bar g\, a^{\dag}, \qquad
e^{s \hat n}\, a\, e^{-s \hat n} = e^{-s}\, a, \qquad
e^{\bar g a^{\dag}}a e^{-\bar g a^{\dag}} = a-\bar g.
\ea
Comparing (\ref{A.3}) and (\ref{A.2}) we get 
\be
\mu \hat n+A a+\bar A a^{\dag} = f'+\bar g' a^{\dag}+s'(\hat n-\bar g a^{\dag})+g' e^{-s}(a-\bar g),
\ee
that gives the equations
\be
\mu = s', \qquad A = e^{-s}g', \qquad \bar A = \bar g'-\bar g s', \qquad f'-e^{-s}\bar g g' = 0.
\ee
to be solved with $s(0) = g(0) = \bar g(0) = f(0) = 0$. The solution is 
\ba
s(\tau)=\mu \tau, \qquad g(\tau) = A\frac{e^{\mu\tau}-1}{\mu}, \qquad 
\bar g(\tau) = \bar A\frac{e^{\mu\tau}-1}{\mu}, \qquad f(\tau) = |A|^{2}\frac{e^{\mu\tau}-1-\mu\tau}{\mu^{2}} .
\ea
%\be
%s(\tau)=\mu \tau
%\ee
%\be
%g'=Ae^{\mu\tau}\qquad\to\qquad g(\tau) = A\frac{e^{\mu\tau}-1}{\mu}
%\ee
%\be
%\bar g' = \bar A+\mu\bar g\qquad\to\qquad \bar g(\tau) = \bar A\frac{e^{\mu\tau}-1}{\mu}
%\ee
%\be
%f' = e^{-\mu\tau}\bar g e^{\mu\tau}A = A\bar g = |A|^{2}\frac{e^{\mu\tau}-1}{\mu}, \qquad\to\qquad f(\tau) = |A|^{2}\frac{e^{\mu\tau}-1-\mu\tau}{\mu^{2}}.
%\ee
Thus, using the explicit $f(\tau)$ at $\tau=1$ in (\ref{3.71}), we get (\ref{3.72}).

\subsection{Direct evaluation by coherent states methods}

The factorization (\ref{A.1}), while general, can be bypassed  for the specific  ground state matrix element
\be
\mmm{0}{e^{\mu \hat n+A\,a+\bar A\,a^{\dag}}}{0}.
\ee
In fact, we can use a more direct evaluation by using coherent state methods. We start by shifting creation and annihilation
operators  by a c-number (leaving their algebra unchanged) 
\be
a = b-\frac{\bar A}{\mu}, \qquad a^{\dag}=b^{\dag}-\frac{A}{\mu}.
\ee
This gives 
\be
\la{A.10}
\mu a^{\dag}a+A a+\bar A a^{\dag} = \mu b^{\dag}b-\frac{|A|^{2}}{\mu} .
\ee
Now, the key remark is that 
\be
b\ket{0} = \zeta\ket{0},\qquad \zeta = \frac{\bar A}{\mu},
\ee
showing that the vacuum $\ket{0}$ is a coherent state $\ket{\zeta}$ for $b, b^{\dag}$, where in general, see Appendix \ref{app:coherent},
\be
\la{A.12}
\ket{\zeta} = e^{-\frac{1}{2}|\zeta|^{2}}\sum_{n=0}^{\infty}\frac{\zeta^{n}}{\sqrt{n!}}\ket{n}, \qquad   \zeta\in\mathbb C .
\ee
Using (\ref{A.12}) we get 
\ba
\mmm{\zeta}{e^{\mu b^{\dag}b}}{ \zeta} &= e^{-|\zeta|^{2}}\sum_{n,m}\frac{\bar \zeta^{n}}{\sqrt{n!}}\frac{\zeta^{m}}{\sqrt{m!}}
\mmm{n}{e^{\mu b^{\dag}b}}{m} = e^{-|\zeta|^{2}}\sum_{n}\frac{|\zeta|^{2n}}{n!}e^{\mu n} \lp
= e^{-|\zeta|^{2}+|\zeta|^{2}e^{\mu}}
\ea
Thus, from (\ref{A.10}), we obtain 
\be
\mmm{0}{e^{\mu \hat n+A\,a+\bar A\,a^{\dag}}}{0} = e^{|\zeta|^{2}(e^{\mu}-1)-\frac{|A|^{2}}{\mu}} = 
e^{\frac{|A|^{2}}{\mu^{2}}(e^{\mu}-1)-\frac{|A|^{2}}{\mu}} 
=e^{\frac{|A|^{2}}{\mu^{2}}(e^{\mu}-\mu-1)},
\ee
which is indeed (\ref{3.72}).

\section{Details of Krylov correlators for the harmonic oscillator}
\la{app:HO}

Let us give full details of the harmonic oscillator complexities with various initial states, \cf \ref{3.92}.

\subsection{Initial TFD state}

Let us follow \cite{Balasubramanian:2022tpr} and consider the harmonic oscillator  
with spectrum $E_{n}=\omega n$ (no zero-point energy). 
If the initial state is the TFD state, the partition function
and the survival amplitude are given by, \cf (\ref{2.10}) and (\ref{2.12}),
\be
Z_{\beta} = \sum_{n=0}^{\infty}e^{-\beta\, n\omega} = \frac{1}{1-e^{-\beta\omega}}\qquad \to\qquad
S(t) = \frac{1-e^{-\beta\omega}}{1-e^{-(\beta-it)\omega}}. 
\ee
The Lanczos coefficients associated with the moments computed from $S(t)$ are
\ba
\la{B.2}
a_{n} &= n\,\frac{\omega}{\tanh(\beta\omega/2)}+\frac{\omega}{e^{\beta\omega}-1}, \qquad
b_{n} = n\,\frac{\omega}{2\sinh(\beta\omega/2)},
\ea
as one can check from the moment recursion (\ref{2.5},\ref{2.6}). \footnote{
A more direct proof is possible based on the $SU(1,1)$ squeezed state structure
\be
\ket{\psi_{\beta}} = \frac{1}{\sqrt{Z_{\beta}}}\sum_{n}e^{-\frac{1}{2}n\beta\omega}\ket{n}_{L}\otimes \ket{n}_{R} = 
\frac{1}{\sqrt{Z_{\beta}}}\sum_{n}e^{-\frac{1}{2}n\beta\omega}\frac{1}{n!}(a^{\dag}_{L})^{n}(a^{\dag}_{R})^{n}\ket{0,0}
= \frac{1}{\sqrt{Z_{\beta}}}\exp\bigg(e^{-\frac{1}{2}\beta\omega}a^{\dag}_{L}a^{\dag}_{R}\bigg)\ket{0,0}.
\ee
}
Comparing this with the Lanczos coefficients in (\ref{3.7}) we get the $\mathfrak{sl}(2,\mathbb R)$ dictionary
\be
\la{B.4}
h = \frac{1}{2}, \qquad \gamma = \frac{\omega}{\tanh(\beta\omega/2)}, \qquad \alpha = \frac{\omega}{2\sinh(\beta\omega/2)}, \qquad \delta = -\frac{1}{2}\omega.
\ee
and notice the combination
\be
4\alpha^{2}-\gamma^{2} = -\omega^{2}.
\ee
Substituting (\ref{B.4}) into (\ref{3.15}) we get
%\ba
%C_{0} +C_{0}^{*} &=  -2\log\bigg[\frac{1}{2}\frac{\cosh(\omega\beta)-\cos(\omega t)}{\sinh^{2}\frac{\beta\omega}{2}}\bigg], \qquad
%|C_{-1}|^{2} = \frac{2\sin^{2}\frac{\omega t}{2}}{\cosh(\omega\beta)-\cos(\omega t)}.
%\ea
%Using also $h=1/2$ in the general formula (\ref{3.13}) we get 
\be
\la{B.6}
C(t) = \frac{\sin^{2}\frac{\omega t}{2}}{\sinh^{2}\frac{\omega \beta}{2}}.
\ee
The expressions of the higher 1-point functions are, \cf (\ref{3.32}),
\bea
C^{(2)}(t) &= \bigg(1+2\frac{\sin^{2}\frac{\omega t}{2}}{\sinh^{2}\frac{\omega\beta}{2}}\bigg)\, \frac{\sin^{2}\frac{\omega t}{2}}{\sinh^{2}\frac{\omega \beta}{2}},\\
C^{(3)}(t) &= \bigg(1+6\frac{\sin^{2}\frac{\omega t}{2}}{\sinh^{2}\frac{\omega\beta}{2}}+6\frac{\sin^{4}\frac{\omega t}{2}}{\sinh^{4}\frac{\omega\beta}{2}}\bigg)\,
 \frac{\sin^{2}\frac{\omega t}{2}}{\sinh^{2}\frac{\omega \beta}{2}}.
\eea
The antisymmetrized 2- and 3-point functions are, \cf (\ref{3.45}),
\bea
A_{2}(t_{1},t_{2}) &= -i\, \frac{\sinh(\omega\beta)}{\sinh^{4}\frac{\beta\omega}{2}}\sin\frac{\omega t_{1}}{2}\sin\frac{\omega t_{12}}{2}\sin\frac{\omega t_{2}}{2}, \\
A_{3}(t_{1},t_{2},t_{3}) &= i\, \frac{\sinh(\omega\beta)}{\sinh^{4}\frac{\beta\omega}{2}}\sin\frac{\omega t_{12}}{2}\sin\frac{\omega t_{13}}{2}\sin\frac{\omega t_{23}}{2},
\eea
where we recall the notation $t_{ij}=t_{i}-t_{j}$.

\subsection{Initial coherent state}
\la{app:coherent}

The Krylov chain associated with a harmonic oscillator coherent state is an example of 
a Heisenberg ($\mathfrak{h}$) model in the sense of Section~\ref{sec:heis}.
We introduce  ladder operators
\be
\la{B.9}
a = \sqrt\frac{\omega}{2}(x+\frac{i}{\omega}p), \qquad a^{\dag}=\sqrt\frac{\omega}{2}(x-\frac{i}{\omega}p), \qquad [a, a^{\dag}]=1,
\ee
and occupation number states
\be
\ket{n} = \frac{(a^{\dag})^{n}}{\sqrt{n!}}\ket{0}, \qquad a\ket{n}=\sqrt{n}\ket{n-1}, \qquad a^{\dag}\ket{n} = \sqrt{n+1}\ket{n+1}.
\ee
Coherent states are 
\be
\ket{z} = e^{-\frac{1}{2}|z|^{2}}\sum_{n=0}^{\infty}\frac{z^{n}}{\sqrt{n!}}\ket{n}, \qquad z\in\mathbb C,
\ee
and they obey
\be
a\ket{z} = e^{-\frac{1}{2}|z|^{2}}\sum_{n=1}^{\infty}\frac{z^{n}}{\sqrt{n!}}\sqrt{n}\ket{n-1} = z\, \ket{z}.
\ee
For $H=\omega(a^{\dag}a+\frac{1}{2})$ the temporal evolution is 
\ba
a(t) &= e^{itH}a e^{-itH}, \qquad \dot a(t) = i[H,a] = i\omega[a^{\dag}a, a] = -i\omega a(t), \qquad a(t) = e^{-it\omega}a.
\ea
Following \cite{Caputa:2024sux}, we notice that 
\be
\ket{z} = D(z)\ket{0}, \qquad D(z) = e^{z a^{\dag}-\bar z a} = e^{z a^{\dag}}e^{-\bar z a}e^{\frac{1}{2}|z|^{2}},\qquad D(z)^{\dag} = D(z)^{-1},
\ee
and we have
%Using
%\ba
%e^{A a^{\dag}+B a} &= e^{\frac{1}{2}AB}e^{A a^{\dag}}e^{B a} = e^{-\frac{1}{2}AB}e^{Ba}e^{A a^{\dag}},\qquad
%e^{-X}a e^{X} = a -[X,a]+\cdots,
%\ea
%we have
\ba
D(z)^{\dag} a D(z) &= e^{-z a^{\dag}+\bar z a}a e^{z a^{\dag}-\bar z a} = a+[a,z a^{\dag}-\bar z a] = a+z, \\
D(z)^{\dag} a^{\dag} D(z) &=a^{\dag}+\bar z.
\ea
We then have 
\be
\wt H = D(z)^{\dagger}H D(z) = \omega\bigg[(a^{\dag}+\bar z)(a+z)+\frac{1}{2}\bigg] = \omega\bigg(a^{\dag}a+\frac{1}{2}+|z|^{2}+\bar z\, a+z\, a^{\dag}\bigg),
\ee
where the tilde here denotes conjugation by $D(z)$, unrelated to the tilde notation used elsewhere.
The action of $\wt H$ in the occupation number basis is tridiagonal
\ba
\wt H\ket{n} &= \omega\bigg(|z|^{2}+\frac{1}{2}+n\bigg)\ket{n}+\omega\bar z\sqrt{n}\ket{n-1}+\omega z\sqrt{n+1}\ket{n+1}.
\ea
Setting
\be
\ket{\wt K_{n}} = e^{in\arg z}\ket{n},
\ee
this is 
\ba
\wt H\ket{\wt K_{n}} &= \omega\bigg(|z|^{2}+\frac{1}{2}+n\bigg)\ket{\wt K_{n}}+\omega|z|\sqrt{n}\ket{\wt K_{n-1}}+\omega|z|\sqrt{n+1}\ket{\wt K_{n+1}},
\ea
that corresponds to the real Lanczos coefficients
\be
a_{n} = \omega\bigg(|z|^{2}+\frac{1}{2}+n\bigg), \qquad b_{n}=\omega|z|\sqrt{n}.
\ee
Comparing with (\ref{3.65}), we have the simple relation 
\be
\l = \omega|z|.
\ee
The one-point complexities are then, \cf (\ref{3.69}), (\ref{3.75}) and (\ref{3.74}), see \cite{Balasubramanian:2022tpr} for the standard complexity, 
\bea
C(t) &=  4|z|^{2}\sin^{2}\frac{\omega t}{2}, \\
C^{(2)}(t) &= 4|z|^{2}\bigg(1+4|z|^{2}\sin^{2}\frac{\omega t}{2}\bigg)\sin^{2}\frac{\omega t}{2}, \\
C^{(3)}(t) &= 4|z|^{2}\bigg(1+12|z|^{2}\sin^{2}\frac{\omega t}{2}+16|z|^{4}\sin^{4}\frac{\omega t}{2}\bigg)\sin^{2}\frac{\omega t}{2}.
\eea
The 2-point and 3-point antisymmetrized functions are, \cf (\ref{3.83}, \ref{3.84})
\bea
A_{2}(t_{1},t_{2}) &=  -8i|z|^{2}\sin\frac{\omega t_{1}}{2}\sin\frac{\omega(t_{1}-t_{2})}{2}\sin\frac{\omega t_{2}}{2},\\
A_{3}(t_{1},t_{2},t_{3}) &=8i|z|^{2}\sin\frac{\omega t_{12}}{2}\sin\frac{\omega t_{13}}{2}\sin\frac{\omega t_{23}}{2}.
\eea

\subsection{Initial squeezed state}

We consider an initial state of the form
\be
\ket{K_{0}} = S(\xi) \ket{0}, \qquad S(\xi) = \exp\bigg[\frac{1}{2}(\bar \xi\, a^{2}-\xi\, a^{\dag 2})\bigg], \qquad \xi\in\mathbb C.
\ee
For real $\xi$ one has 
\be
\la{B.26}
S(\xi)\ket{x} = e^{-\xi/2}\ket{e^{-\xi}x},
\ee
and thus 
\be
K_{0}(x) = \mmm{x}{S(\xi)}{0} = e^{-\xi/2}\psi_{0}(e^{-\xi}x),
\ee
which is a Gaussian state with generic width. 
For the unit mass oscillator we then consider
\be
K_{0}(x) =  \bigg(\frac{1}{2\pi r}\bigg)^{1/4}e^{-\frac{x^{2}}{4r}},
\ee
where $r$ is a generic real parameter. The ground state is $r = 1/(2\omega)$.
As shown in \cite{Hashimoto:2023swv}, the full Krylov chain  is independent of $\omega$ and reads
\be
K_{n}(x) = \braket{x}{K_{n}} = \frac{(-1)^{n}}{(2\pi r)^{1/4}}\frac{1}{\sqrt{(2n)!}2^{n}}H_{2n}\bigg(\frac{1}{\sqrt{2r}}x\bigg)\, e^{-\frac{x^{2}}{4r}}.
\ee
This gives
\be
a_{n} = \frac{1+4r^{2}\omega^{2}}{2r}\bigg(n+\frac{1}{4}\bigg), \qquad b_{n} = \frac{1-4r^{2}\omega^{2}}{4r}\sqrt{n\bigg(n-\frac{1}{2}\bigg)}.
\ee
Comparing with (\ref{3.7}) gives the identification
\be
h = \frac{1}{4}, \qquad \gamma = \frac{1+4r^{2}\omega^{2}}{2r}, \qquad \alpha = \frac{1-4r^{2}\omega^{2}}{4r}.
\ee
The fact that we get a $\mk{sl}(2,\mathbb R)$ case is related to the oscillator representation of $\mk{sl}(2,\mathbb R)$ in terms of $a^{2}, a^{\dag 2}$ and $H$. This explains the 
special value $h=1/4$. 

The one-point complexities are then, \cf (\ref{3.32}),
\bea
C(t) &=  \frac{(1-4r^{2}\omega^{2})^{2}}{32r^{2}\omega^{2}}\sin^{2}(\omega t), \\
C^{(2)}(t) &= \frac{(1-4r^{2}\omega^{2})^{2}}{32r^{2}\omega^{2}}\bigg(1+\frac{3(1-4r^{2}\omega^{2})^{2}}{32r^{2}\omega^{2}}\sin^{2}(\omega t)\bigg)\, \sin^{2}(\omega t),  \\
C^{(3)}(t) &= \frac{(1-4r^{2}\omega^{2})^{2}}{32r^{2}\omega^{2}}\bigg(1+\frac{9(1-4r^{2}\omega^{2})^{2}}{32r^{2}\omega^{2}}\sin^{2}(\omega t)
+\frac{15(1-4r^{2}\omega^{2})^{4}}{1024 r^{4}\omega^{4}}\sin^{4}(\omega t)\bigg)\, \sin^{2}(\omega t).
\eea
The 2-point and 3-point antisymmetrized functions are, \cf (\ref{3.83}, \ref{3.84}),
\bea
A_{2}(t_{1},t_{2}) &=  i\frac{(1-4r^{2}\omega^{2})^{2}(1+4r^{2}\omega^{2})}{64r^{3}\omega^{3}}\sin(\omega t_{1})\sin(\omega t_{12})\sin(\omega t_{2}), \\
A_{3}(t_{1},t_{2},t_{3}) &= i\frac{(1-4r^{2}\omega^{2})^{2}(1+4r^{2}\omega^{2})}{64r^{3}\omega^{3}}\sin(\omega t_{12})\sin(\omega t_{13})\sin(\omega t_{23}).
\eea

 \bibliography{Krylov-Biblio}
\bibliographystyle{JHEP-v2.9}
\end{document}